\documentclass[reprint, showpacs,
preprintnumbers,
amsmath,amssymb,
aps, floatfix,
superscriptaddress,
prd, nofootinbib]{revtex4-1}

\usepackage{float}
\usepackage{graphics}% Include figure files
\usepackage{dcolumn}% Align table columns on decimal point
\usepackage{tikz}
\usepackage{bm}% bold math
\usepackage{hyperref}% add hypertext capabilities
\hypersetup{colorlinks=false,
            citebordercolor=1 1 1,
            linkbordercolor=1 1 1,
            pdfborderstyle={/S/N},
            pdfborder=0 0 0}
\usepackage{xcolor}
\usepackage{import}
\usepackage{booktabs}
\usepackage{multirow}
\usepackage[capitalise, nameinlink]{cleveref}
\usepackage{makecell}
\usepackage{apptools}
\usepackage{subcaption}
\usepackage{ragged2e} % for justified captions
\usepackage{enumerate}% http://ctan.org/pkg/enumerate
\usepackage{orcidlink}
\usepackage{soul}

\newcommand{\RevA}[1]{{\color{black}#1}}

\usepackage{adjustbox}

\begin{document}

\title{Measurements and models of enhanced recombination \\following inner-shell vacancies in liquid xenon}% Force line breaks with \\

%\title{LZ Author List for Paper (RevTeX4 Draft 241030T1759(GMT))}
%\date{October 30, 2024 17:59 (GMT)}

% 1 
\author{J.~Aalbers}
\affiliation{SLAC National Accelerator Laboratory, Menlo Park, CA 94025-7015, USA}
\affiliation{Kavli Institute for Particle Astrophysics and Cosmology, Stanford University, Stanford, CA  94305-4085 USA}

% 2 
\author{D.S.~Akerib}
\affiliation{SLAC National Accelerator Laboratory, Menlo Park, CA 94025-7015, USA}
\affiliation{Kavli Institute for Particle Astrophysics and Cosmology, Stanford University, Stanford, CA  94305-4085 USA}

% 3 
\author{A.K.~Al Musalhi \orcidlink{0009-0004-3067-9939}}
\email{aiham.almusalhi@ucl.ac.uk}
\affiliation{University College London (UCL), Department of Physics and Astronomy, London WC1E 6BT, UK}

% 4 
\author{F.~Alder}
\affiliation{University College London (UCL), Department of Physics and Astronomy, London WC1E 6BT, UK}

% 5 
\author{C.S.~Amarasinghe \orcidlink{0000-0001-7789-651X}}
\email{chami@ucsb.edu}
% 6 
\affiliation{University of California, Santa Barbara, Department of Physics, Santa Barbara, CA 93106-9530, USA}

% 7 
\author{A.~Ames}
\affiliation{SLAC National Accelerator Laboratory, Menlo Park, CA 94025-7015, USA}
\affiliation{Kavli Institute for Particle Astrophysics and Cosmology, Stanford University, Stanford, CA  94305-4085 USA}

% 8 
\author{T.J.~Anderson}
\affiliation{SLAC National Accelerator Laboratory, Menlo Park, CA 94025-7015, USA}
\affiliation{Kavli Institute for Particle Astrophysics and Cosmology, Stanford University, Stanford, CA  94305-4085 USA}

% 9 
\author{N.~Angelides}
\affiliation{Imperial College London, Physics Department, Blackett Laboratory, London SW7 2AZ, UK}

% 10 
\author{H.M.~Ara\'{u}jo}
\affiliation{Imperial College London, Physics Department, Blackett Laboratory, London SW7 2AZ, UK}

% 11 
\author{J.E.~Armstrong}
\affiliation{University of Maryland, Department of Physics, College Park, MD 20742-4111, USA}

% 12 
% ...

\collaboration{The LZ Collaboration}

\date{\today}

\begin{abstract}
Electron-capture decays of $^{125}$Xe and $^{127}$Xe, and double-electron-capture decays of $^{124}$Xe, are backgrounds in searches for weakly interacting massive particles (WIMPs) conducted by dual-phase xenon time projection chambers such as LUX-ZEPLIN (LZ).
These decays produce signals with more light and less charge than equivalent-energy $\beta$ decays, and correspondingly overlap more with WIMP signals.
We measure three electron-capture charge yields in LZ: the 1.1~keV M-shell, 5.2~keV L-shell, and 33.2~keV K-shell at drift fields of 193~V/cm and 96.5~V/cm.
The LL double-electron-capture decay of $^{124}$Xe exhibits even more pronounced shifts in charge and light.  
We provide a first model of double-electron-capture charge yields using the link between ionization density and electron-ion recombination, and identify a need for more accurate calculations.
Finally, we discuss the implications of the reduced charge yield of these decays and other interactions creating inner-shell vacancies for future dark matter searches. 

\end{abstract}

%\keywords{Suggested keywords}%Use showkeys class option if keyword
                              %display desired
\maketitle
\newcommand{\xefour}{$^{124}$Xe}
\newcommand{\xefive}{$^{125}$Xe}
\newcommand{\xeseven}{$^{127}$Xe}
\newcommand{\sy}{_\text{sys}}

\newcommand{\HL}[1]{\textcolor{green}{HL: #1}}
\section{\label{sec:introduction}Introduction}

The direct search for galactic dark matter in the form of weakly interacting massive particles (WIMPs) is currently led by three experiments using dual-phase xenon time projection chambers (TPCs)~\cite{aalbers2023first, bo2024dark, aprile2023first,aalbers2024dark}. 
A major factor in the success of this detector technology is the inherent ability to distinguish between WIMP-like nuclear recoil (NR) events and background-like electron recoil (ER) events by measuring the charge-to-light ratio in the detector response.
For a given combined charge and light signal, ER interactions exhibit a relatively higher charge yield than NR events~\cite{akerib2020discrimination}. 

Recent observations from the LUX and XELDA experiments suggest that inner-shell vacancies (ISVs) created by electron-capture (EC) decays produce less charge and more light than $\beta$ decays, which are typically used to calibrate the ER response in dark matter experiments~\cite{akerib2017ultralow,temples2021measurement}.
With their lower charge yields, EC decays appear more NR-like than their $\beta$ decay counterparts.
Further, ISVs created by the double-electron-capture (DEC) decay of $^{124}$Xe -- which has the longest measured half-life of any known decay -- are now also observed~\cite{xenon2019observation, aprile2022double, aalbers2024two}, with more charge suppression than EC decays.

Here, we report measurements of the charge yields of $^{125}$Xe and $^{127}$Xe EC decays in LUX-ZEPLIN (LZ), which clearly exhibit enhanced recombination.
As both decays come with associated $\gamma$-ray emission, we employ two complementary techniques: a two-vertex selection as performed in LUX~\cite{akerib2017ultralow}, and a selection where the $\gamma$ ray escapes the TPC as in XELDA~\cite{temples2021measurement}. 
We then argue that $^{124}$Xe DEC decays exhibit additional charge suppression relative to EC decays, as observed in LZ's latest WIMP analysis~\cite{aalbers2024dark}. 
We propose increased recombination as an explanation for this effect, parametrized using the Thomas-Imel box (TIB) model~\cite{thomas1987recombination, szydagis2review}.

ISVs created by DEC decays and neutrino-electron scattering are irreducible backgrounds with rates that scale linearly with dark matter search exposure.
We find that the WIMP search data are statistically inconsistent with DEC decay models based on naive extrapolations of $\beta$ decay and EC charge yields, even at current exposures [4], but are consistent with an extrapolation of EC charge yields motivated by the TIB model.
Despite this, we find that uncertainties in the DEC charge yield model do not significantly impact LZ's WIMP sensitivity.
We argue that better measurements and models will be needed for candidates whose signals overlap significantly with the DEC background and next-generation experiments~\cite{aalbers2022next,gaspert2022neutrino,temples2021measurement}. 

\subsection{\label{subsec:micro}Signal production in liquid xenon}

Energy deposited by particles scattering in liquid xenon (LXe) excites and ionizes atoms, producing observable light and charge, with some energy lost to atomic motion.
Excited atoms interact with neighboring atoms to form excimers, which relax with decay time constants of $\lesssim 30$~ns by emitting vacuum ultraviolet (VUV) photons at 175~nm~\cite{kubota1978evidence, fujii2015high}.
Additional VUV photons are produced via delayed excimer channels when a fraction of ionization electrons recombine with ions~\cite{chepel2013liquid}.
The fraction of electrons escaping recombination may be increased by applying a stronger electric field, which correspondingly decreases the amount of light produced.\footnote{The conversion between electrons and photons through recombination is assumed to be one-to-one, though slight departures from this have been observed~\cite{anton2020measurement}.}
 
In ER interactions, most of the energy is deposited through electronic stopping, resulting in a total number of observed quanta proportional to the deposited energy, but in NR events, most of the deposited energy is lost to nuclear stopping, generating atomic motion (heat) that creates no visible signal in the TPC.
This difference leads to two distinct scales for reconstructed energy:  ER-equivalent (keV$_\text{ee}$) energy, which is linear in the number of quanta observed, and NR-equivalent energy (keV$_\text{nr}$), which is a nonlinear function of the number of quanta observed. 
For instance, a 25~keV$_\text{nr}$ NR event produces approximately the same number of observable quanta as a 5~keV$_\text{ee}$ ER event, so 25~keV$_\text{nr} \sim $ 5~keV$_\text{ee}$.
In the remainder of the article, we exclusively use the~keV$_\text{ee}$ scale (and~keV$_\text{ee}$ label) for reconstructed energy, while labeling known true energies with~keV.
We also make references to the WIMP search region, which for LZ is $\lesssim$~15~keV$_\text{ee}$.

The basis for distinguishing ER events from NR events in LXe hinges on the differing numbers of electrons -- relative to the total quanta produced --  ionized by each type of interaction. 
ER events create more electron-ion pairs per quanta produced than NR events. 
This distinction is characterized through the exciton-ion ratio \(N_\text{ex}/N_\text{i}\) which, in the search region for WIMPs, stands at less than 0.1 for ER events~\cite{szydagis2review} and exceeds 0.7 for NR events~\cite{Dahl:2009nta}.

After the initial partitioning of energy into excitons and ions, the fraction $r$ of ions that recombine determines the final charge and light yields.
We focus on the charge yields of the ER sources discussed in this work, as dual-phase TPCs such as LZ (described in Sec.~\ref{subsec:lz}) detect charge with a much higher efficiency than light. 
The charge yield $Q_y$ is defined as the number of nonrecombined electrons $(1-r)N_\text{i}$ per unit energy, such that
\begin{equation}\label{eq:qy_full}
    Q_y = \frac{(1-r)N_\text{i}}{W(N_\text{i} + N_\text{ex})} = \frac{1-r}{W \left( 1 + N_\text{ex}/N_\text{i} \right)},
\end{equation}
where $W$ is the average energy required to generate a single electron or VUV photon.
We assume a $W$-value of $13.5$~eV, consistent with values adopted in WIMP searches~\cite{aprile2023first, aalbers2024dark, meng2021dark}, but note that two recent experiments have observed an average of $W=11.5\pm0.6$~eV~\cite{anton2020measurement, baudis2021measurement}, a discrepancy that would impact the $Q_y$ measurements in this work. 
Therefore, where possible, we also report charge yield ratios that are insensitive to the $W$-value.
The exciton-ion ratio for ER events is modeled with energy dependence given in Ref.~\cite{szydagis2review} such that it goes to zero at low energy, and varies between $0.01$ and $0.1$ in the WIMP search region.

Following excitation, ionization, and recombination, ER interactions produce more observable charge (and less light) than NR interactions with the same number of observable quanta, enabling discrimination. 
Benchmarks in dual-phase TPCs based on a 50\% acceptance of various WIMP hypotheses show leakage from a flat-in-energy ER background into WIMP detection regions at levels ranging from 0.5\% to 0.03\%~\cite{szydagis2review}.
Although our analysis highlights shifts in the average ER recombination fraction, variance in $r$ and detector resolution effects also impact ER leakage~\cite{akerib2020discrimination}.

\begin{figure}[t]
    \includegraphics[width=1\linewidth]{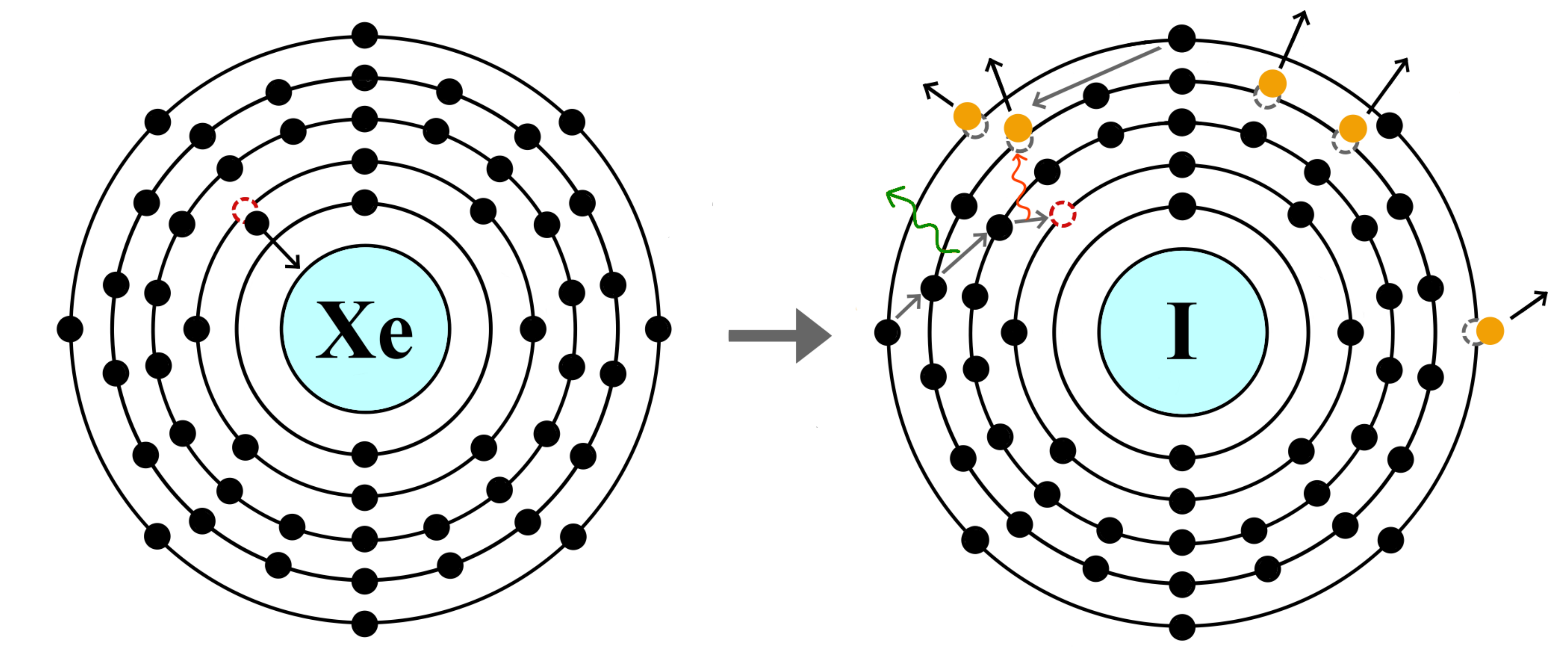}
    \caption{\justifying \small Simplified schematic of xenon decaying to iodine via electron capture (left), and the ISV relaxing (right) by emitting {\color{black} Auger-Meitner} electrons (yellow) and X-rays. 
    A virtual photon is shown in red to illustrate the Auger-Meitner process, as well as an X-ray in green.}
    \label{fig:EC_scheme}
\end{figure}

We attribute the reduced charge yield observed in EC decays to a higher $r$. 
This fraction can be influenced by a variety of factors such as the strength of the electric field, electron diffusion, recombination cross-section, and particularly the ionization density, which is sensitive to the topology of energy deposition~\cite{szydagis2011nest, szydagis2013enhancement}. 
An EC decay process is illustrated in Fig.~\ref{fig:EC_scheme} to highlight how the total energy of the decay is distributed among multiple low-energy {\color{black} Auger-Meitner} electrons. 
This increased electron multiplicity and the lower energy of individual electrons collectively result in a denser ionization cloud than that produced by a single-electron track of equivalent energy. 
This phenomenon is discussed further using the Thomas-Imel box model~\cite{thomas1987recombination} in Section~\ref{subsec:box_model}.

\section{\label{sec:methods}Methods}

\subsection{\label{subsec:lz}The LZ experiment}

LZ is a dark matter direct detection experiment at the Sanford Underground Research Facility (SURF) in Lead, South Dakota, USA. 
It is located 4850~ft underground in the Davis Cavern, where the rock overburden provides 4300~m of water-equivalent shielding. 
The core of the LZ detector is a cylindrical dual-phase xenon TPC with a 7~tonne active LXe volume.
The TPC is embedded in two veto detectors designed to tag multi-site interactions that form a background to the WIMP search.
Immediately enveloping the TPC is the LXe ``Skin" designed to tag $\gamma$ rays, and surrounding the xenon cryostat is the near-hermetic outer detector (OD) filled with gadolinium-loaded liquid scintillator, designed to tag neutrons. 
The entire assembly is submerged in ultra-pure water for further shielding. 
More details on the LZ detector design and assembly may be found in Refs.~\cite{LZ_NIM, LZ_TDR}.

VUV photons and electrons are both detected in the TPC, which is instrumented with arrays of photomultiplier tubes (PMTs) on the top and bottom. 
Following an energy deposition, prompt VUV photons form the primary scintillation signal (S1). Ionization electrons drift upward under the influence of an electric field and are subsequently extracted into a gaseous xenon region by a stronger field, where a secondary scintillation signal (S2) is formed via electroluminescence. 
Both electric fields are relevant to our measurements, as the drift field affects $r$ and the extraction field determines the gain of the S2 channel.

In addition to ER-NR discrimination, the S1 and S2 signals also enable position reconstruction of individual scatters. 
The transverse position is derived from the S2 hit pattern on the top PMT array, and the depth of the interaction is inferred from the time separation of the S1 and S2 signals.
The number of S2 signals contained within a given event window determines whether the event type is a multiple scatter (MS) or single scatter (SS), the two relevant classes in this analysis.

Radioactive sources are used as standard candles for signal correction and to calibrate the detector response.
S1 and S2 signal sizes are corrected for the position-dependent light collection efficiency, the depth-dependent depletion of drifting electrons as they are captured by electronegative impurities, and time-varying PMT gains. 
The corrected quantities, labeled as S1$c$ and S2$c$, are used hereafter.
The detector response is captured by the quantities $g_1$ and $g_2$: the gains of the S1 and S2 channels, respectively.
The energy deposited by an SS event, consisting of a single S1-S2 pair, is thus reconstructed as 
\begin{equation}\label{eq:e_recon}
    E_\text{ER} = W \left( \frac{\text{S1}c}{g_1} + \frac{\text{S2}c}{g_2} \right).
\end{equation}
\\ \\
In this analysis, the measured charge yields are defined as
\begin{equation}\label{eq:qy_recon}
    Q_y = \frac{\text{S2}c}{g_2 E},
\end{equation}
where we divide by the true energy $E$ instead of the reconstructed energy $E_\text{ER}$ of Eq.~\ref{eq:e_recon}.

Two LZ datasets are used: the first science run (WS2022)~\cite{aalbers2023first} and the second, longer run (WS2024)~\cite{aalbers2024dark}, with the corresponding drift field and gains summarized in Table~\ref{tab:run}.
LZ employs the NEST package to simulate signal production in the detector~\cite{szydagis2011nest, szydagis2review}. 
The default ER yield model from NEST~v2.4.0 matches the WS2022 tritium $\beta$ calibration {\color{black} within statistical uncertainty, and is therefore used unchanged in WS2022.
For WS2024, the NEST model was fine-tuned to match the higher-statistics tritium calibration taken at the start of the run.
This fine-tuning was performed in S1$c$--S2$c$ space using a Markov chain Monte Carlo algorithm~\cite{foreman2013emcee, foreman2019emcee}.
The modeling error is better than 0.2\% in mean log$_{10}$S2$c$ -- calculated in 2~phd S1$c$ bins -- over the WIMP search energy range, and is negligible for the purposes of this analysis.}
We use ``$\beta$ model'' to refer to these NEST models for the remainder of this work. 
Further details of the calibrations and corrections in each run may be found in Refs.~\cite{aalbers2023first, aalbers2024dark}.

\begin{table}
\caption{\justifying \label{tab:run}% 
\small Drift field and detector gains in the two LZ science runs; the WS2022 value of $g_2$ slightly differs from that in Ref.~\cite{aalbers2023first} due to updated correction maps.
The units of gains are in photons detected (phd) per quantum.
}
\begin{ruledtabular}
\begin{tabular}{lccc}
    \textrm{Run} & \textrm{Drift field [V/cm]} & \textrm{$g_1$ [phd/photon]} & \textrm{$g_2$ [phd/$e^-$]} \\
    \colrule \noalign{\vskip 1mm}
    WS2022 & 193 & 0.113 (3) & 48.9 (7)\\
    WS2024 & 96.5 & 0.112 (2) & 34.0 (9) \\
\end{tabular}
\end{ruledtabular}
\end{table}

\subsection{\label{sec:source_description}Description of EC and DEC sources}

The EC sources used for the charge yield measurements are $^{127}$Xe ($t_{1/2} = 36.4$~d) and $^{125}$Xe ($t_{1/2} = 16.9$~h)~\cite{TabRad_v8}.
The WS2022 measurement relies on $^{127}$Xe produced via cosmogenic activation before the xenon was brought underground. The cosmogenic activity was depleted by the start of WS2024, and hence the $^{125}$Xe and $^{127}$Xe used that run are primarily from activation during neutron calibrations{\color{black}, with significantly lower event rates than in WS2022.}
For the purpose of this work, it is assumed that measurements are not sensitive to subtle differences between the de-excitation cascades of the two isotopes, and they are hence taken to be identical.
We observe $^{124}$Xe, a source of DEC decays with a natural abundance of $0.0952 \pm 0.0003$\% and a half life of $t_{1/2} = 1.1 \times 10^{22}$~yr~\cite{valkiers1998primary, aprile2022double, aalbers2024two}, creating 6.3 events per tonne-year of exposure in the WIMP region. 

We use the \{K, L, M, N\} International Union of Pure and Applied Chemists notation to refer to the principal electron shells ($n = 1, 2, 3, 4$), while sub-shells (such as L$_1$ and L$_2$) indicate specific energy levels within these shells corresponding to different $\ell$ and $j$ quantum numbers. 

\subsubsection{EC on $^{125}\mathrm{Xe}$ and $^{127}\mathrm{Xe}$}
\label{subsec:ECdesc}

When a $^{127}$Xe ($^{125}$Xe) nucleus captures an atomic electron and forms an excited $^{127}$I ($^{125}$I) nucleus, a higher-shell electron fills the ISV and sets off an atomic cascade.
The energy differences between cascading shells are released as X-rays or Auger-Meitner electrons until the binding energy of the orbital hole is expended.
The most important $^{127}$Xe EC probabilities and iodine binding energies are listed in Table~\ref{tab:ec_capture_prob}, and are assumed to be the same for $^{125}$Xe. 
In addition to the atomic de-excitation, the excited iodine nucleus relaxes to the ground state via internal conversion (IC) electron or $\gamma$-ray emission. 
Figures~\ref{fig:Xe127decay} and \ref{fig:Xe125decay} depict the nuclear decay schemes of $^{127}$Xe and $^{125}$Xe, respectively.
\begin{table}
\caption{\justifying \label{tab:ec_capture_prob}%
\small Atomic sub-shell energies of iodine, and associated probabilities of EC in xenon~\cite{booklet2001x, mougeot2019towards} (only the most relevant sub-shells for EC are listed). 
{\color{black}We use the L$_1$ and M$_1$ energies (corresponding to $\ell=0$) throughout this analysis. 
} 
}
\begin{ruledtabular}
\begin{tabular}{ccc}
    \textrm{Subshell} & \textrm{Energy [keV]} & \textrm{Capture probability [\%]} \\
    \colrule \noalign{\vskip 1mm}
    K$_1$ & 33.1694 & 84.40 (3) \\
    L$_1$ & 5.1881 & 12.011 (17) \\
    L$_2$ & 4.8521 & 0.3375 (5) \\
    M$_1$ & 1.0721 & 2.444 (10) \\
    M$_2$ & 0.9305 & 0.07168 (17) \\
    N$_1$ & 0.1864 & 0.609 (5) \\
    N$_2$ & 0.1301 & 0.01697 (12) \\
    O$_1$ & 0.0136 & 0.1100 (17) \\
    O$_2$ & 0.0038 & 0.00197 (3) \\
\end{tabular}
\end{ruledtabular}
\end{table}

L-capture is primarily followed by pure {\color{black} Auger-Meitner} emission, with X-ray fluorescence emitted in only 9\% of decays~\cite{nuclearcharts}.
Between seven and nine {\color{black} Auger-Meitner} electrons are emitted following an L-capture with average energy below 1~keV: seven are expected under the hypothesis that one {\color{black} Auger-Meitner} electron induces the emission of two higher-energy electrons~\cite{nagy2009radiochemistry}, while xenon ions with +8 and +9 charge have been experimentally observed~\cite{mohammedeinFinalCharge}.
The track length of a 5.2~keV electron in LXe is around 100~nm while for a 1~keV electron it is 10~nm~\cite{berger2002estar}, leading to $O(10^3)$ differences in the volumetric track density between L-captures and 5.2~keV $\beta$ decays.

In 88\% of K-captures, a 28.3--28.6~keV X-ray is emitted, leaving an L vacancy that relaxes via {\color{black} Auger-Meitner} cascade~\cite{nuclearcharts}.
The X-ray will travel $O(100)$~µm in LXe before depositing its energy~\cite{hubbell1995tables}, which is too short of a distance to resolve in the LZ detector. Therefore, the entire event is observed as an SS. From a recombination standpoint, however, the post-capture atomic emission is a multi-site event, because $O(100)$~µm is far beyond the recombination range of the primary site~\cite{mozumder1995free}; two independent sets of light and charge quanta are generated.

\definecolor{charcoal}{rgb}{0.21, 0.27, 0.31}
\begin{figure}[t]
\begin{adjustbox}{center}
\begin{subfigure}{\textwidth}
    \begin{tikzpicture}[
    scale=0.9,
    level2/.style={very thick},
    level/.style={thick},
    trans/.style={thick,->,shorten >=1pt,>=stealth},
    frac/.style={very thin, dashed, charcoal}
    ]
    % Draw the energy levels.
    \draw[level2] (1cm,11.5em) -- (2cm,11.5em) node[midway,above] {\large{\textbf{$^{127}$Xe}}};
    % \draw[level] (2.8cm,10em) -- (8.2cm,10em);
    % \node[] at (7.9,4.1){618};
    \draw[level] (2.8cm,10em) -- (8.2cm,10em);
    \node[] at (7.9,3.5){375};
    \draw[level] (2.8cm,5.4em) -- (8.2cm,5.4em);
    \node[] at (7.9,2.0){202.8};
    \draw[level] (2.8cm,2em) -- (8.2cm,2em);
    \node[] at (7.9,0.9){57.6};
    \draw[level2] (2.8cm,-1em) -- (8.2cm,-1em) node[midway, below] {\large{\textbf{$^{127}$I}}};
    % Draw the transitions.
    \draw[frac] (1.5cm, 11.5em)-- (1.5cm,5em)node[midway, above, rotate=90]{\scriptsize{36.4 d}};
    % \draw[frac] (1.5cm, 10em)-- (2.8cm,10em)node[midway, above]{\scriptsize{0.0143\%}};
    \draw[frac] (1.5cm, 10em)-- (2.8cm,10em)node[midway, above]{\scriptsize{47.6\%}};
    \draw[frac] (1.5cm, 5.4em)-- (2.8cm,5.4em) node[midway, above]{\scriptsize{53\%}};
    % \draw[trans] (3.6cm,10em) -- (3.6cm,-6em);
    % \node[rotate=45] at (3.7,11em){\footnotesize{\textbf{618}}};
    % \node[rotate=90] at (3.3,2.5){\footnotesize{0.0143\%}};
    \draw[trans] (3.3cm,10em) -- (3.3cm,-1em);
    \node[rotate=45] at (3.4,11em){\footnotesize{\textbf{375}}};
    \node[rotate=90] at (3,2.5){\footnotesize{17.3\%}};
    \draw[trans] (4.25cm,10em) -- (4.25cm,5.4em);
    \node[rotate=45] at (4.35,11.3em){\footnotesize{\textbf{172.1}}};
    \node[rotate=90] at (3.95,2.5){\footnotesize{25.7\%}};
    \draw[trans] (5.2cm,5.4em) -- (5.2cm,-1em);
    \node[rotate=45] at (5.3,6.8em){\footnotesize{\textbf{202.8}}};
    \node[rotate=90] at (4.9,1.2){\footnotesize{68.7\%}};
    \draw[trans] (6.15cm,5.4em) -- (6.15cm,2em);
    \node[rotate=45] at (6.25,6.8em){\footnotesize{\textbf{145.3}}};
    \node[rotate=90] at (5.85,1.2){\footnotesize{4.31\%}};
    \draw[trans] (7.1cm,2em) -- (7.1cm,-1em);
    \node[rotate=45] at (7.2,3.2em){\footnotesize{\textbf{57.6}}};
    \node[rotate=90] at (6.8,0.15){\footnotesize{1.24\%}};
    \end{tikzpicture}
    \subcaption{}
    \label{fig:Xe127decay}
\end{subfigure}
\end{adjustbox}
\bigskip
\begin{adjustbox}{center}
\begin{subfigure}{\textwidth}
\begin{tikzpicture}[
    scale=0.9,
    level2/.style={very thick},
    level/.style={thick},
    trans/.style={thick,->,shorten >=1pt,>=stealth},
    frac/.style={very thin, dashed, charcoal}
    ]
    % Draw the energy levels.
    \draw[level2] (1cm,11.5em) -- (2cm,11.5em) node[midway,above] {\large{\textbf{$^{125}$Xe}}};
    \draw[level] (2.8cm,10em) -- (8.2cm,10em);
    \node[] at (7.9,3.5){1089};
    \draw[level] (2.8cm,6em) -- (8.2cm,6em);
    \node[] at (7.9,2.2){453.8};
    \draw[level] (2.8cm,2em) -- (8.2cm,2em);
    \node[] at (7.9,0.9){243.4};
    \draw[level] (2.8cm,-1em) -- (8.2cm,-1em);
    \node[] at (7.9,-0.1){188.4};
    \draw[level2] (2.8cm,-4em) -- (8.2cm,-4em) node[midway, below] {\large{\textbf{$^{125}$I}}};
    % Draw the transitions.
    \draw[frac] (1.5cm, 11.5em)-- (1.5cm,-1em)node[midway, above, rotate=90]{\scriptsize{16.9 h}};
    \draw[frac] (1.5cm, 10em)-- (2.8cm,10em)node[midway, above]{\scriptsize{1.9\%}};
    \draw[frac] (1.5cm, 6em)-- (2.8cm,6em)node[midway, above]{\scriptsize{4.7\%}};
    \draw[frac] (1.5cm, 2em)-- (2.8cm,2em) node[midway, above]{\scriptsize{66.6\%}};
    \draw[trans] (3.3cm,10em) -- (3.3cm,2em);
    \draw[frac] (1.5cm, -1em)-- (2.8cm,-1em) node[midway, above]{\scriptsize{25.4\%}};
    \node[rotate=45] at (3.4,11.5em){\footnotesize{\textbf{846.5}}};
    \node[rotate=90] at (3,2.6){\footnotesize{1.11\%}};
    \draw[trans] (4.25cm,6em) -- (4.25cm,-4em);
    \node[rotate=45] at (4.35,7.6em){\footnotesize{\textbf{453.8}}};
    \node[rotate=90] at (3.95,1.3){\footnotesize{4.7\%}};
    \draw[trans] (5.2cm,2em) -- (5.2cm,-4em);
    \node[rotate=45] at (5.3,3.4em){\footnotesize{\textbf{243.4}}};
    \node[rotate=90] at (4.9,-0.8){\footnotesize{30\%}};
    \draw[trans] (6.15cm,2em) -- (6.15cm,-1em);
    \node[rotate=45] at (6.25,3.2em){\footnotesize{\textbf{55.0}}};
    \node[rotate=90] at (5.85,0.15){\footnotesize{6.8\%}};
    \draw[trans] (7.1cm,-1em) -- (7.1cm,-4em);
    \node[rotate=45] at (7.2,0.4em){\footnotesize{\textbf{188.4}}};
    \node[rotate=90] at (6.8,-0.8){\footnotesize{53.8\%}};
    \end{tikzpicture}
     \subcaption{}
     \label{fig:Xe125decay}
\end{subfigure}
\end{adjustbox}
\bigskip
\begin{adjustbox}{center}
\begin{subfigure}{\textwidth}
\begin{tikzpicture}[
    scale=0.9,
    level2/.style={very thick},
    level/.style={thick},
    trans/.style={thick,->,shorten >=1pt,>=stealth},
    frac/.style={very thin, dashed, charcoal}
    ]
    % Draw the energy levels.
    \draw[level2] (1cm,11.5em) -- (2cm,11.5em) node[midway,above] {\large{\textbf{$^{124}$Xe}}};
    \draw[level2] (2.8cm,5em) -- (8.2cm,5em) node[midway, below] {\large{\textbf{$^{124}$Te}}};
    % Draw the transitions.
    \draw[frac] (1.5cm, 11.5em)-- (1.5cm,5em)node[midway, above, rotate=90]{\scriptsize{$1.8\times10^{22}$ y}};
    % \draw[frac] (1.5cm, 10em)-- (2.8cm,10em)node[midway, above]{\scriptsize{0.0143\%}};
    \draw[frac] (1.5cm, 5em)-- (2.8cm,5em) node[midway, above]{\scriptsize{$\approx$100\%}};
    \end{tikzpicture}
     \subcaption{}
     \label{fig:Xe124decay}
\end{subfigure}
\end{adjustbox}
\caption{\justifying Nuclear decay scheme of $^{127}$Xe (a),  $^{125}$Xe (b) and $^{124}$Xe (c) showing only states with a branching ratio $>1\%$~\cite{branchingratios}. 
The number above each transition is the $\gamma$-ray energy in keV, while that to the side is the percentage of parent decays that involve the transition.
Relaxation through IC emission is not included in the indicated $\gamma$ intensities.
The bold horizontal lines represent the ground states of the respective nuclei, while the finer ones mark excited states of the iodine isotopes, with their energy in keV on the right. 
}
\end{figure}
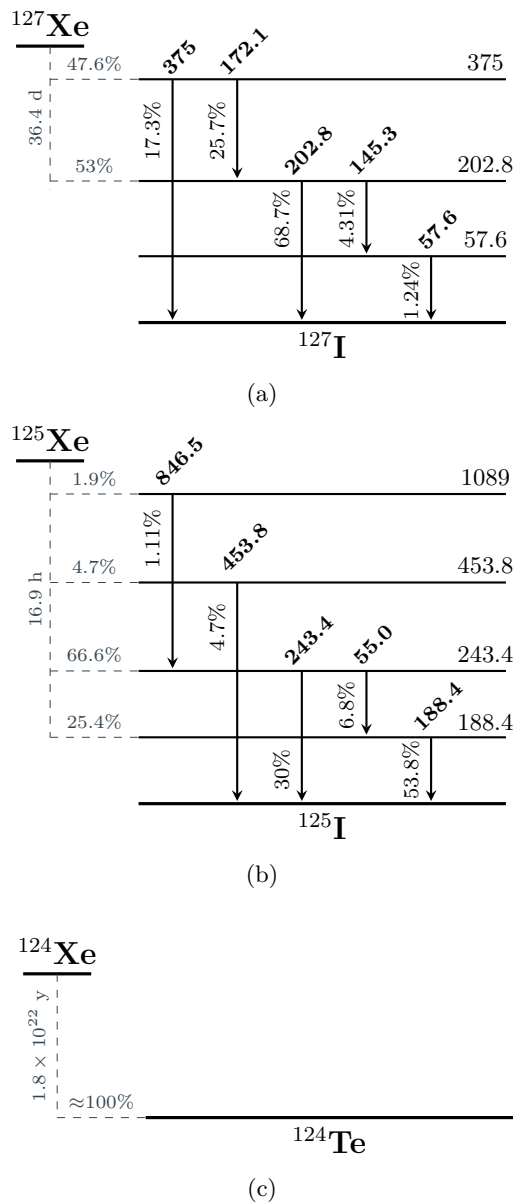

\subsubsection{DEC on $^{124}\mathrm{Xe}$}

The simultaneous capture of two electrons by a $^{124}$Xe nucleus to form a ground state $^{124}$Te nucleus~\cite{COELLOPEREZ2019134885} launches an atomic cascade similar to that in EC. 
%not produce an excited daughter nucleus but rather one already at the ground state
Since $^{124}$Te is produced already in the ground state, there are no associated $\gamma$ rays, and $^{124}$Xe DEC decays manifest entirely as SS events throughout the bulk of the LXe.
The \mbox{L$_1$L$_1$-}, L$_1$M$_1$-, L$_1$N$_1$-, and M$_1$M$_1$-captures (written henceforth without subscripts) of $^{124}$Xe are of special interest as they fall directly within the energy range for a WIMP search. 
The $^{124}$Xe DEC probabilities and energies are listed in Table~\ref{tab:dec_counts}, taken from recent calculations~\cite{nictescu2024theoretical} that included MM-captures for the first time~\cite{priv_comm}.

\begin{table}
    \caption{\justifying \small Theoretical atomic relaxation energies and DEC probabilities for $^{124}$Xe~\cite{nictescu2024theoretical}. 
    These numbers differ slightly from those used in the background model of the WS2024 dark matter search~\cite{aalbers2024dark}, but were chosen for this work because the calculation in Ref.~\cite{nictescu2024theoretical} also generated predictions for the LN and MM decays~\cite{priv_comm}. 
    }
    \begin{ruledtabular}
        \begin{tabular}{lcc}
         Subshells & Energy [keV] & Capture probability [\%] \\
        \colrule \noalign{\vskip 1mm}
         KK & 64.62 & 74.13--74.15 \\
         KL$_1$ & 37.05 & 18.76--18.83 \\
         KM$_1$ & 32.98 & 3.83--3.84 \\
         KN$_1$ & 32.11 & 0.83--0.85 \\
         KO$_1$ & 31.93 & 0.13 \\
         L$_1$L$_1$ & 10.04 & 1.22 \\
         L$_1$M$_1$ & 6.01 & 0.49 \\
         L$_1$N$_1$ & 5.37 & 0.27 \\
         M$_1$M$_1$ & 2.05 & 0.13 \\
    \end{tabular}
    \end{ruledtabular}
    \label{tab:dec_counts}
\end{table}

\subsection{EC event selections}

\subsubsection{\label{subsec:multiple_EC_measurements_MS}MS selection of K-, L-, and M-captures}

\begin{figure}[t]
\begin{subfigure}{0.75\columnwidth}
\includegraphics[width=\columnwidth]{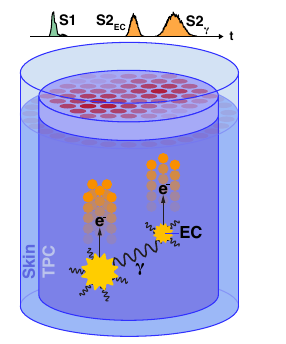}
\caption{} \label{fig:diagram_MS}
\end{subfigure}
\bigskip
\begin{subfigure}{0.75\columnwidth}
\includegraphics[width=\columnwidth]{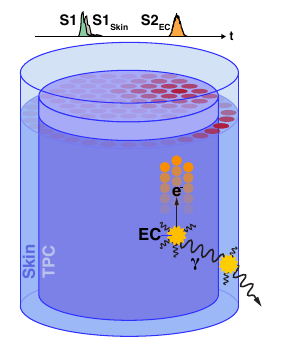}
\caption{} \label{fig:diagram_SS}
\end{subfigure}
\caption{\justifying \small Schematics of EC decay events with either an MS (a) or Skin-tagged SS (b) topology, along with waveforms.
The MS signature has one merged S1 and two S2s: one from the decay site and one from the vertically displaced $\gamma$-ray site.  
The SS case has an S1--S2 pair solely from the atomic cascade, with the outgoing $\gamma$-ray interaction generating a pulse in the Skin coincident with the TPC S1.
} \label{fig:TPC_Diagram}
\end{figure}

The MS selection keeps $^{127}$Xe and $^{125}$Xe EC events  where the associated nuclear $\gamma$ rays travel far enough such that the $\gamma$ photoabsorption and atomic cascade sites are spatially resolved.
The resulting events present a single, summed S1 and multiple S2s: one from the atomic cascade and one for each separately resolved $\gamma$ ray. 
We further select decays with a single transition to the ground state, which reduces the multiplicity of S2s to two: one from the atomic cascade and one from the $\gamma$-ray interaction. 
This choice also excludes events involving Compton scattering nuclear $\gamma$ rays, which would yield multiple S2s.
The event topology is illustrated in Fig.~\ref{fig:diagram_MS}.
For this analysis, we select for the 203~keV and 243~keV $\gamma$ rays for $^{127}$Xe and  $^{125}$Xe, respectively.
Higher energy transitions are excluded to avoid selecting $^{214}$Pb decays to the 290~keV excited state of $^{214}$Bi, which present the same MS topology.
Decays to the 188~keV excited state of iodine have a lower efficiency for reconstruction as MS events due to the shorter scattering lengths of the resulting $\gamma$ rays, and are not specifically targeted in the MS selection.

A clear separation of the two S2 pulses in drift time is crucial to the charge yield measurement of each deposition.
When the $\gamma$ ray is emitted perpendicularly to the drift field, its ionization electrons will reach the liquid surface at the same time as the electrons from the atomic cascade.
This creates a merged S2 that hinders the MS classification and biases the charge yield measurement. 
On the other hand, if the $\gamma$ ray interacts above or below the decay site, a time delay between the drifting electron clouds is introduced, making their S2 signals more resolvable.
We impose a minimum separation of 3.5~µs in drift time between the two pulses to reliably resolve them, which corresponds to an average minimum vertical distance of 0.8~cm between the energy depositions.
Furthermore, events where the atomic cascade is observed second (is lower in the TPC) are discarded to avoid contamination from secondary electrons emitted as a result of the larger S2 produced by the $\gamma$-ray interaction.

Further cuts are applied to increase the purity of the selection.
Events with a coincident signal in the Skin detector are removed.
Accidental coincidences of SS events with isolated S2s in the detector are mitigated by i) only keeping events within 61~cm of the central TPC axis to avoid external $\gamma$ rays, and ii) cutting events where the scatter vertices are separated by more than 5~cm {\color{black} in three dimensions} -- informed by the roughly 1~cm mean-free-path of a 250~keV $\gamma$ ray~\cite{gaikwad2017mass}.
The top and bottom of the TPC, and regions close to the TPC field-cage resistors are excluded to mitigate elevated radioactivity.
This results in an estimated fiducial mass of 4.35~tonnes for this analysis in both runs.
Finally, the total reconstructed energy is restricted to a 195--290~keV range to select the relevant excited states of the iodine nuclei. 

For K-shell events, an additional cut is required to suppress contamination from neutron-activated $^{129\text{m}}$Xe.\footnote{This paragraph, Tab.~\ref{tab:measurements}, Fig.~\ref{fig:dists}, and Fig.~\ref{fig:ec_means} have been updated following publication of an erratum~\cite{erratum} to the originally published article~\cite{prd_version}. The main conclusions of the article are unchanged.}
These decays involve two nuclear transitions at 40~keV and 197~keV, producing a total deposited energy that overlaps with the $^{127}$Xe K-shell peak (203 + 33.2~keV). 
While the relaxations typically emit localized conversion electrons, in 5\% of decays the 197~keV transition produces a $\gamma$ ray that can become spatially separated from the primary vertex, generating an MS topology.
For the WS2022 dataset, where cosmogenic $^{127}$Xe dominates over $^{129\text{m}}$Xe, temporal cuts are applied to remove periods of elevated $^{129\text{m}}$Xe activity. 
Specifically, the first 16 days of the science run and the 16 days following neutron calibrations are excluded.
For WS2024 however, the diminished abundance of $^{127}$Xe and the higher frequency of neutron calibrations make this approach impractical.
Instead, the K-shell analysis is restricted to $^{125}$Xe decays by requiring minimum total deposited energy of 260~keV.

\begin{figure}[t]
    \centering
    \includegraphics[width=0.95\linewidth]{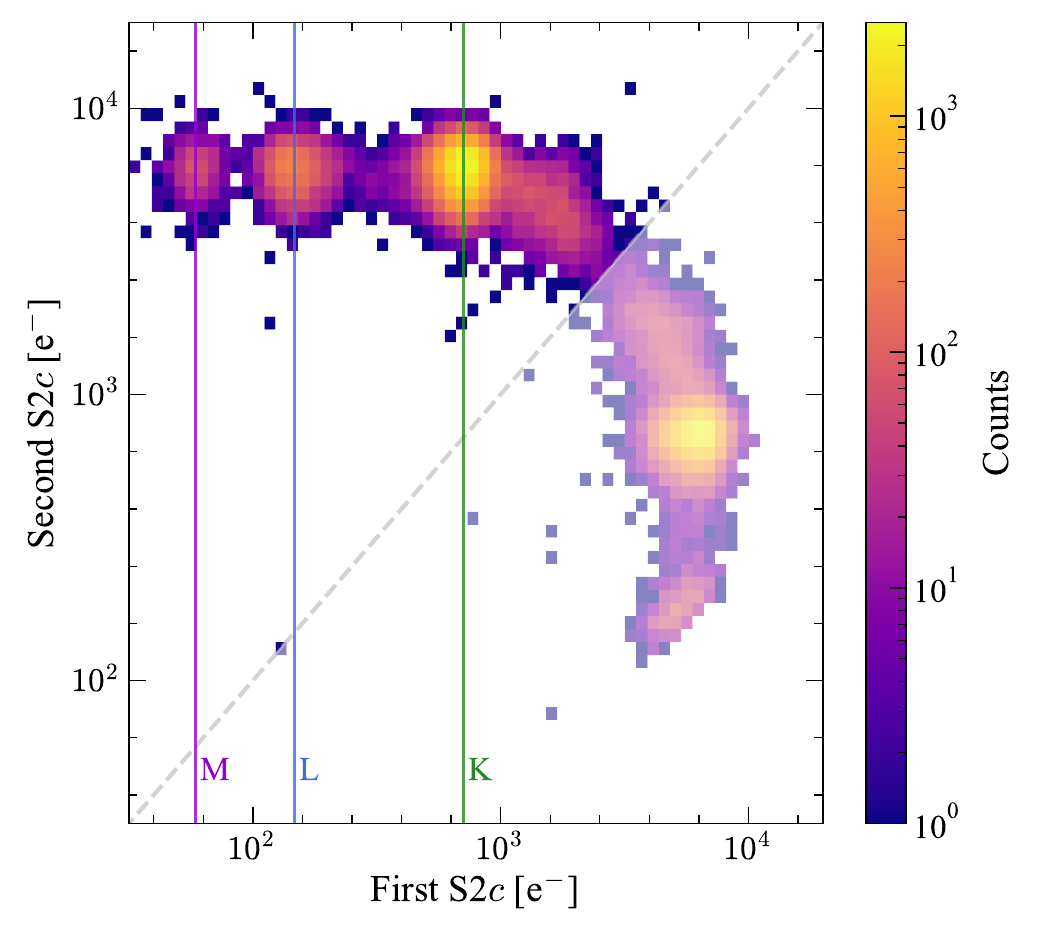}
    \caption{\justifying \small Selection of $^{127}$Xe EC decays in WS2022 ($^{125}$Xe is subdominant in this run). The corrected area of the second S2 to reach the liquid surface is plotted against that of the first. S2 areas are in units of electrons; the single electron size in WS2022 is 58.5~phd~\cite{aalbers2023first}. The horizontal arm (above the diagonal line) identifies events in which the S2 of the atomic cascade arrives first, i.e. the where the atomic cascade occurs above the nuclear $\gamma$ energy deposit. 
    Events where the S2 of the $\gamma$ deposit arrives first (below the diagonal line) are not used in this analysis since the smaller, second S2 from the atomic cascade is obscured by the tail of the first.
    The continuum of events adjacent to the diagonal is formed by the two-step decay from the 203~keV state, where the S2 from the 58~keV $\gamma$~ray merges with the EC.
    }
    \label{fig:selection_2022}
\end{figure}

Figure~\ref{fig:selection_2022} showcases the paired S2$c$ distributions in the WS2022 dataset, with clearly distinguishable M-, \mbox{L-}, and K-capture populations.
The diagonal cut restricts the selection to events where the atomic cascade happens above the $\gamma$ site.

\subsubsection{\label{subsec:single_EC_measurements_SS}SS selection of K- and L-captures}

As illustrated in Fig.~\ref{fig:diagram_SS}, EC decays near the edges of the TPC can be selected by looking for events where the associated $\gamma$ ray escapes but leaves a coincident pulse in the Skin detector.
This approach presents a trade-off: some separation  distance between the EC decay and the TPC wall is required to mitigate localized field distortions near the wall, but too much separation relative to the mean free path of the $\gamma$ ray decreases the signal efficiency.
The separation was chosen to be 1~cm and 1.5~cm for WS2022 and WS2024 data, respectively, by identifying the point at which the energy reconstruction and resolution of 33.2~keV K-shell EC events and  163.9~keV $^{131\mathrm{m}}$Xe IC events deviate from those within the LXe bulk.
The integrity of the position-dependent corrections was further verified using tritium calibration data.

\begin{figure}[t]
     \centering
     \includegraphics[width=1.0\linewidth]{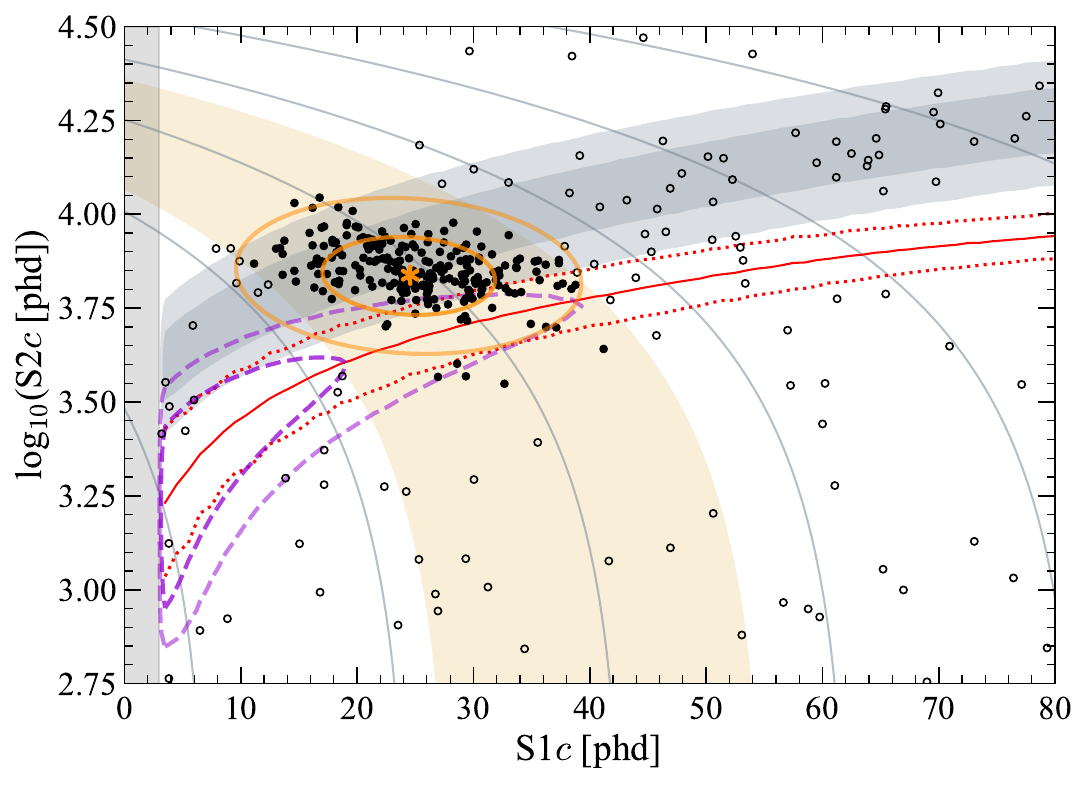}
     \caption{\justifying \small Skin-tagged events in WS2022 data within the expanded volume optimized for the SS EC analysis.  Orange curves show the 1 and 2$\sigma$ contours of a Gaussian fit to the L-capture peak.  The fit is based on the population of solid points, defined by an energy selection (tan shading) and a loose 4$\sigma$ cut around the $\beta$-decay background region (1 and 2$\sigma$ regions indicated by gray bands).  The hollow points outside the fit window include a population of low-S2 events that are absent in the WIMP search fiducial volume and are expected from field non-uniformities and charge loss near the TPC wall.  There is a distinct shift in the L-capture population down from the $\beta$-decay band, encroaching on the WIMP region of interest, shown by the dashed purple lines (1 and 2$\sigma$ contours for a 30 GeV/c$^2$ WIMP) and red lines (centroid and 1$\sigma$ contours for a flat-in-energy NR signal).}
     \label{fig:Lshell_S1S2}
 \end{figure}

A given event is prompt-tagged by the Skin veto if a pulse within the Skin detector both exceeds 3~phd in area and occurs within 500~ns of the S1 pulse in the TPC.
Skin-tagged events exhibit clear peaks at the K-shell and L-shell energies, and a further selection is made by restricting the reconstructed energy in the TPC to within $2\sigma$ of the reconstructed peak means.
Additional cuts are applied to remove mis-reconstructed events and emissions from TPC field-cage resistors and electrode grids.
{\color{black} A 34.5~cm separation was set from the top and bottom of the active volume to exclude regions of high Skin-tagging activity.}
The WS2022 L-shell SS selection is depicted in Fig.~\ref{fig:Lshell_S1S2} to demonstrate the visible shift of ECs with respect to ER interactions down into the WIMP signal region, highlighting the importance of accurately modeling recombination enhancements.
While this method allows isolation of the S1 from the atomic cascade, it is limited in statistics compared to the MS approach and does not retain a significant sample of M or higher shell captures.

\section{\label{sec:results}Results}

\begin{figure*}
    \centering
    \includegraphics[width=1
\linewidth]{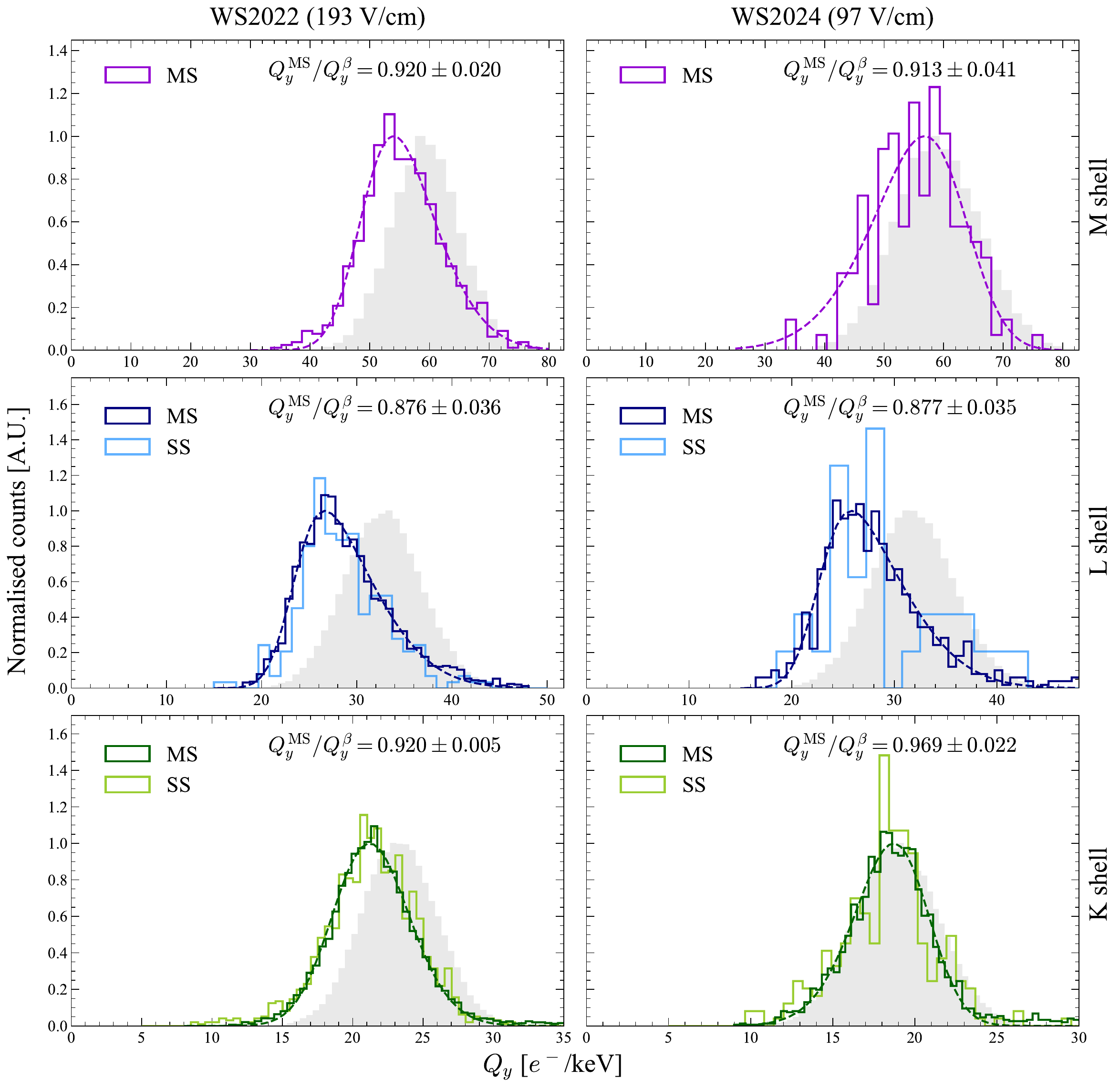}
    \caption{\justifying \small The full set of measured EC charge yield distributions in LZ data at two different field configurations, for each of K-, L-, and M-shell captures. With the exception of the M-shell case, Skin-tagged (SS) samples are overlaid onto MS selections to highlight their compatibility. Skew-Gaussian fits to the MS distributions are also shown as dashed lines. To demarcate where ECs lie with respect to standard ER interactions, the yields from equivalent mono-energetic electrons drawn from the NEST $\beta$ model are included in gray.}
    \label{fig:dists}
\end{figure*}

\begin{figure}
    \centering
    \includegraphics[trim={0.2cm 0cm 0cm 0cm},clip,width=0.93\linewidth]{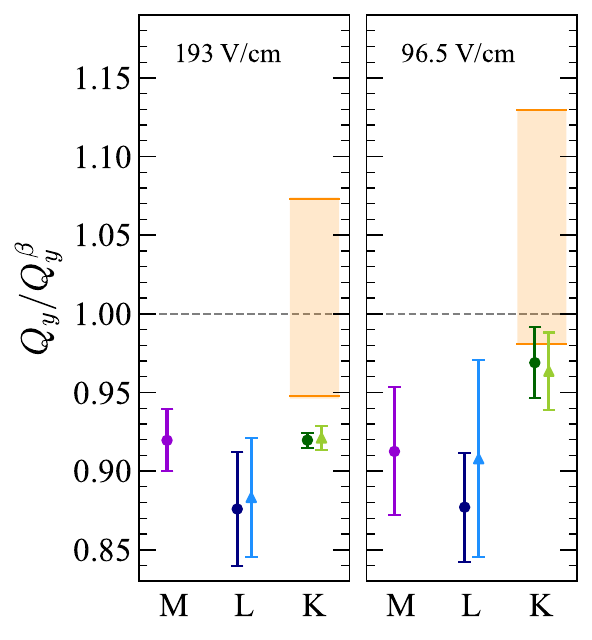}
    \caption{\justifying \small Summary of charge yield ratios measured with the MS selection (dark circles) and the SS selection (light triangles). 
    The M- and L-capture $Q_y$ ratios should be compared directly with the $\beta$ model (dashed line), while the K-capture measurements should be compared to the multi-site expectation (orange bands), where the lower (upper) line corresponds to X-ray modeled in NEST as a $\gamma$ ($\beta$) interaction. 
    }
    \label{fig:ec_means}
\end{figure}

Charge yield distributions for M-, L-, and K-capture events using MS and SS selections are displayed in Fig.~\ref{fig:dists}, and compared to simulations of mono-energetic $\beta$ decays at each corresponding energy.
The EC distributions are fitted with skew-Gaussian functions -- shown for the MS selection in Fig.~\ref{fig:dists} -- and the mean absolute charge yields $Q_y^\text{EC}$ [$e^-$/keV] extracted from the fits are listed in Table~\ref{tab:measurements}.
Charge yield ratios $Q_y^\text{EC}/Q_y^{\beta}$ are also reported alongside the absolute charge yields and summarized in Fig.~\ref{fig:ec_means}.

The energies used in the denominators of $Q_y^\text{EC}$ and $Q_y^\beta$ are the M$_1$ and L$_1$ iodine binding energies from Table~\ref{tab:ec_capture_prob} instead of the reconstructed energies from the SS selection; the MS selection cannot reconstruct the energy of the atomic cascade separately from the $\gamma$ ray as their S1s are merged.
The simulated $Q_y^\beta$ reference yields are obtained by evaluating the tuned LZ $\beta$ model at the true EC energies.

\addtolength{\tabcolsep}{+10pt} % Add space between columns 
\begin{table*}
\caption{\justifying 
\small Summary of the EC charge yield and ratio measurements in both LZ runs, LUX~\cite{akerib2017ultralow}, and XELDA~\cite{temples2021measurement}. 
In calculating the ratios, the default NEST $\beta$ model is used in WS2022, and a custom-tuned NEST model is used in WS2024.
The ratios of the other experiments are calculated using their own $\beta$ calibrations.
}
\centering
\begin{tabular}{ lccc } 
\hline
\hline
\noalign{\vskip 1mm}
Run & Source & $Q_y^\text{EC}$ [$e^-$/keV] & $Q_y^\text{EC}/Q_y^\beta$ \\
\colrule \noalign{\vskip 1mm}
\multirow{5}{*}{LZ WS2022 (193~V/cm)} & M (MS) & $55.75 \pm 0.26_\text{stat} \pm 1.13\sy$ & $0.920 \pm 0.004_\text{stat} \pm 0.019\sy$ \\ 
& L (MS) & $28.68 \pm 0.13_\text{stat} \pm 0.58\sy$ & $0.876 \pm 0.004_\text{stat} \pm 0.036\sy$ \\
& L (SS) & $28.92 \pm 0.38_\text{stat} \pm 0.45\sy$ & $0.883 \pm 0.012_\text{stat} \pm 0.036\sy$\\ 
& K (MS) & $21.43 \pm 0.05_\text{stat} \pm 0.31\sy$ & $0.920 \pm 0.002_\text{stat} \pm 0.004\sy$\\ 
& K (SS) & $21.46 \pm 0.12_\text{stat} \pm 0.30\sy$ & $0.921 \pm 0.005_\text{stat} \pm 0.006\sy$\\ 
\colrule \noalign{\vskip 1mm}
\multirow{5}{*}{LZ WS2024 (96.5~V/cm)} & M (MS) & $54.59 \pm 1.61_\text{stat} \pm 2.49\sy$ & $0.913 \pm 0.027_\text{stat} \pm 0.031_\text{sys}$ \\ 
& L (MS) & $27.81 \pm 0.22_\text{stat} \pm 0.98\sy$ & $0.877 \pm 0.007_\text{stat} \pm 0.034\sy$ \\
& L (SS) & $28.79 \pm 1.76_\text{stat} \pm 0.84\sy$ & $0.908 \pm 0.056_\text{stat} \pm 0.029\sy$\\ 
& K (MS) & $18.35 \pm 0.07_\text{stat} \pm 0.61\sy$ & $0.969 \pm 0.004_\text{stat} \pm 0.022\sy$\\ 
& K (SS) & $18.25 \pm 0.24_\text{stat} \pm 0.48\sy$ & $0.964 \pm 0.013_\text{stat} \pm 0.021\sy$\\ 
\colrule \noalign{\vskip 1mm}
\multirow{4}{*}{LUX (180~V/cm)} & N (MS) & $75.3 \pm 6.5 _\text{stat} \pm 5.2\sy$ & $1.151 \pm 0.099 _\text{stat} \pm 0.080\sy$ \\
& M (MS) & $61.4 \pm 0.5_\text{stat} \pm 4.3\sy$ & $1.127 \pm 0.009 _\text{stat} \pm 0.079\sy$ \\
& L (MS) & $30.8 \pm 0.1_\text{stat} \pm 2.1\sy$ & $0.928 \pm 0.003 _\text{stat} \pm 0.063\sy$ \\
& K (MS) & $22.72 \pm 0.03_\text{stat} \pm 1.58\sy$ & $0.984 \pm 0.001 _\text{stat} \pm 0.068\sy$ \\
\colrule \noalign{\vskip 1mm}
XELDA (258~V/cm) & L (SS) & $32.87 \pm 0.07_\text{stat} \pm 0.37\sy$ & $0.909 \pm 0.003_\text{stat} \pm 0.007\sy$ \\
\colrule \noalign{\vskip 1mm}
XELDA (363~V/cm) & L (SS) & $33.63 \pm 0.03_\text{stat} \pm 0.33\sy$ & $0.917 \pm 0.001_\text{stat} \pm 0.009\sy$ \\
\hline
\hline
\end{tabular}
\label{tab:measurements}
\end{table*}
\addtolength{\tabcolsep}{-10pt}    
 
As described in Sec. \ref{subsec:ECdesc}, K-capture decays are multi-site, featuring a 28.3--28.6~keV X-ray in addition to an L-shell vacancy. For simplicity, we report $Q_y^\text{K}/Q_y^{\beta}$ with respect to a mono-energetic $\beta$ with the full K-shell energy, though we show in Fig.~\ref{fig:ec_means} orange bands that represent expectations when the two sites are treated independently.
There are two calculations used to bound the multi-site comparison in Fig.~\ref{fig:ec_means}: one where the X-ray charge yield is taken from the $\beta$ model, and another where it is taken from an alternative ``$\gamma$ model'' available in NEST~v2.4.0, based on soft X-ray yield measurements albeit not yet tuned on LZ data~\cite{szydagis2review, obodovskii1994scintillation}.
In both cases, the yield of the {\color{black} Auger-Meitner} site is evaluated from the $\beta$ model and summed with the X-ray yield.

The SS and MS methods are in good agreement for all three shells.
The M- and L-capture charge yields are clearly suppressed relative to the equivalent-energy $\beta$ yields.
The K-capture measurement in WS2022 is also suppressed compared to the $\gamma+\beta$ multisite expectation, while the WS2024 measurement is consistent, albeit with a larger statistical error.
%Apart from the WS2024 K-capture decays, the SS and MS shapes are in good agreement. 

Skewness is present in many of the EC distributions, but a consistent trend with energy and drift field in the range explored is not apparent.
We note that the positive skewness of the L-capture distributions could be explained by the X-ray emission in 9\% of L-capture decays.
The 3--5~keV X-rays have a range of a few µm and may form separate recombination sites -- similar to K-capture X-rays -- resulting in higher charge yields and introducing a positive skew in the distribution.
The $\beta$ simulations are well-described with standard Gaussian functions, and no evidence for skewness~\cite{akerib2020discrimination, akerib2020improved, szydagis2021review} was present in the WS2022 and WS2024 ER calibration data.

The systematic uncertainties in the $Q_y^\text{EC}$ measurement are: i) a 1--3\% uncertainty in $g_2$, ii) 1--3\% residual errors introduced by the position-based signal corrections, and iii) a 2--3\% residual variation {\color{black} of charge yields in time that we attribute to} PMT gains recovering from high-rate neutron calibrations.
The uncertainty on $g_2$ (and $g_1$) is constrained by the light-charge anti-correlation of sixteen ER calibration peaks, and the other systematic errors are estimated by varying the selections used in the two analyses.
No uncertainty is assumed on the 13.5~eV $W$-value used in these measurements.

Since both the numerator and denominator of the $Q_y^\text{EC}/Q_y^{\beta}$ ratio scale identically with $g_2$ and $W$-value, the ratio itself is not subject to their uncertainties. 
There is an additional error on the ratio (not present for the absolute yield) stemming from the energy reconstruction: the SS reconstructed energies are lower than the tabulated decay energies in both runs, by up to 5\% in the L-captures and less than 2\% for the other shells. 
This discrepancy cannot be completely explained by the uncertainties on $g_1$ and $g_2$ in Table~\ref{tab:run}.
We therefore evaluate the $\beta$ model at both the tabulated and reconstructed energies and take the difference in the resulting charge ratio as an additional systematic uncertainty. 
 
\section{Discussion}

\subsection{\label{subsec:box_model}The Thomas-Imel box model}

We argue that the increased recombination observed in EC events relative to $\beta$ decays is caused by increased track ionization density.
We utilize the Thomas-Imel box (TIB) model~\cite{thomas1987recombination} to explore this link and calculate an expected $^{124}$Xe LL charge yield, \RevA{which is then compared to the value of $Q_y^\text{LL}/Q_y^\beta = 0.70 \pm 0.04$ measured in the recent LZ WIMP search~\cite{aalbers2024dark}}. 
The TIB model describes the electron-ion recombination fraction $r$ of ER events in terms of a single phenomenological parameter $\xi$, such that
\begin{equation} \label{eq:r_xi}
    r = 1 - \frac{\ln{(1+\xi)}}{\xi}.
\end{equation}
The TIB model assumes that recombination takes place in a box of side-length $2a$ within which charges are uniformly distributed.
This simplifying feature allows $\xi$ to be expressed in terms of a few physical parameters as 
\begin{equation}\label{eq:xi}
    \xi = \frac{N_\text{i} \alpha}{4 a^2 v_d}.
\end{equation}
Here, $N_\text{i}$ represents the initial number of ions, $v_d$ is the electron drift speed, and $\alpha$ denotes the associated recombination coefficient.
With this description, the charge yield $Q_y$ defined in Eq.~\ref{eq:qy_full} can be rewritten in terms of $\xi$ as
\begin{equation} \label{eq:TIB}
    Q_y = \frac{\ln( 1 + \xi)}{W \xi \left( 1 + N_\text{ex}/N_\text{i}\right)}.
\end{equation}

The effective ionization density $\rho = N_\text{i}/(8a^3)$ is related to recombination via the $N_\text{i}/a^2$ factor that appears in Eq.~\ref{eq:xi}.
%Both $N_\text{i}$ and $a$ can change with features of the track. 
While $N_\text{i}$ is approximately linear with energy (up to changes in $N_{\text{ex}}/N_\text{i}$), there is currently no accepted convention in the literature for how $a$ depends on features of the track.
No interpretation of $a$ is provided in Ref.~\cite{thomas1987recombination}, which introduced the TIB model; it is regarded as the electron thermalization distance~\cite{mozumder1995free} in Ref.~\cite{szydagis2review} and as a critical electrostatic length scale in Ref.~\cite{Dahl:2009nta}, but   
neither interpretation would predict that the recombination following an {\color{black} Auger-Meitner} cascade is enhanced when compared to a $\beta$ decay of the same energy.  
We emphasize that a recombination enhancement is explained if $a$ also depends on the size of the ionization tracks, which are markedly smaller in EC interactions.
The spatial dependence of an ionization track due to $dE/dx$ might also impact the recombination fraction, but this effect is not captured by the single length scale $a$ of the TIB model. 

Using the TIB model, we wholly attribute the recombination enhancement to differences in the effective ionization density. 
Assuming that both types of interaction produce the same $N_\text{i}$~\footnote{\RevA{This assumption is made for simplicity in the model. A slightly different $N_i$ may be expected given the weakly energy-dependent $N_{ex}/N_i$ in the NEST $\beta$-model, but the difference in $N_i$ is small, and its effect on $Q_y$ is further suppressed by the dependence of $r$ on $N_i$.  The shift in the modeled $Q_y^{LL}/Q_y^\beta$ when the energy-dependent $N_{ex}/N_i$ is taken into account is only 0.003, an order of magnitude smaller than the measured uncertainty.}}, the effective ionization density for $\beta$ interactions is set by a box size $a_\beta$, while for EC interactions it is set by smaller box sizes $a_\text{L}$ and $a_\text{M}$.

%\RevA{To simplify the calculation, we assume that an Auger-Meitner cascade has the same $N_\text{i}$ and $N_\text{ex} / N_\text{i}$ as an equivalent-energy $\beta$. We believe this assumption is justified because  increasing $N_\text{i}$ also leads to additional recombination, leading to about the same final number of electrons. Doing the full calculation for the LL case results in a shift in our modeled $Q_y^\text{LL}/Q_y^\beta$ ratio of $0.003$, an order of magnitude smaller than the measured uncertainty.}

%The assumption that an {\color{red} Auger-Meitner} cascade has the same $N_\text{i}$ and $N_\text{ex} / N_\text{i}$ as an equivalent-energy $\beta$ event allows for a simplified calculation -- {\color{red} described in the rest of this section -- that doesn't require the details of the Auger-Meitner spectrum. 
%We find that accounting for the energy dependence of $N_\text{i}$ and $N_\text{ex} / N_\text{i}$ has a negligible effect on our charge yield result -- the impact is smaller than what can be resolved within our uncertainties, as additional recombination and the increased number of ions largely cancel each other out.
%}

\begin{figure}
    \includegraphics[width=1.0\linewidth]{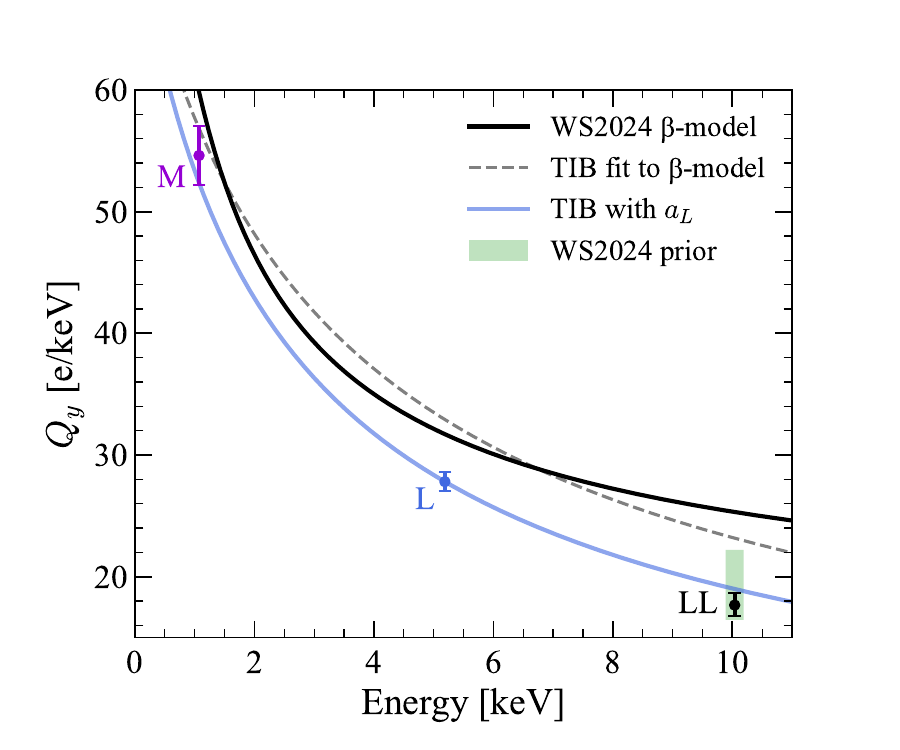}
    \caption{\justifying 
    \small The WS2024 $\beta$ charge yield is shown with the black line. 
    The $\beta$ TIB model, shown in dashed gray, is fit to the $\beta$ yield over the 0.5--11~keV$_\text{ee}$ range, as described in the text.
    The TIB model that matches the L-capture MS measurement (blue point) is shown with the blue line corresponding to a smaller box size. 
    The L-capture TIB model predicts the LL charge yield at $10.04$~keV$_\text{ee}$ and agrees with the best-fit in the WS2024 WIMP search (black point).
    The green rectangle shows the flat prior of the LL charge yield used in the WIMP search fit, and is discussed in the Appendix.
    The M-capture MS measurement is also shown (purple point).
    }
    \label{fig:Qy_E}
\end{figure}

We first fit the LZ WS2024 $\beta$ charge yield curve with the TIB model given in Eqs.~\ref{eq:xi} and \ref{eq:TIB} over the 0.5--11~keV$_\text{ee}$ range in 10.5~eV$_\text{ee}$ intervals.
In each interval, $N_\text{i}$ and $N_\text{ex}/N_\text{i}$ are evaluated with NEST, leaving $4\xi/N_\text{i} = \alpha / (a_\beta^2v_d)$ as the only free parameter in the fit.\footnote{The related parameter $\xi/N_\text{i}$ is called $\varsigma$, e.g., in Ref.~\cite{lenardo2015global}.}
The outcome is drawn as a dashed contour in Fig.~\ref{fig:Qy_E}.
While this implementation of the TIB model does not describe the $\beta$ yields perfectly, it captures the general trend with energy using a single box size.
Next we apply the TIB model to the EC measurements presented in Sec.~\ref{sec:results}. 
For instance, we obtain a value of $\alpha / (a_\text{L}^2 v_d)$ for the WS2024 MS-measured L-capture charge yield by numerically solving Eq.~\ref{eq:TIB}, with $N_\text{i}$ and $N_\text{ex}/N_\text{i}$ evaluated at the L-capture energy using NEST.

Assuming that the recombination coefficient $\alpha$ is the same for EC and $\beta$ events (see below Eq.~15 of Ref.~\cite{szydagis2review}), and since $v_d$ is constant for a given detector configuration, we obtain $a_\text{L}/a_\beta = 0.84 \pm 0.03$ for WS2024, indicating that the effective ionization density in L-capture events is $1.7 \pm 0.2$ times higher than in equivalent-energy $\beta$ events.
These uncertainties are propagated solely from the MS-measurement uncertainty; the WS2024 $\beta$ model describes LZ tritium calibration data to better than 0.2\% in mean log$_{10}$(S2$c$)~\cite{aalbers2024dark}, which is negligible. 
We similarly conclude that the box sizes of the other M- and L-capture events in WS2022 and WS2024 are smaller than the $\beta$  box size; the full list of the relevant TIB model parameters is given in Table~\ref{tab:TIB_results}.

\addtolength{\tabcolsep}{+8pt} % Add space between columns 
\begin{table*}
\caption{\justifying 
\small TIB parameters (Eq.~\ref{eq:TIB}) that reproduce charge yields measured in LZ. 
The values of $N_\text{i}$ and $N_\text{ex}/N_\text{i}$ are obtained from NEST~\cite{szydagis2review} using an LXe density of $2.9$~g/cm$^3$.
Values for the TIB model free parameter $4\xi/N_\text{i} = \alpha/(a^2v_d)$ are calculated as described in the text.
The $N_\text{i}$ and $N_\text{ex}/N_\text{i}$ values for relevant DEC decays are also shown for completeness, with the best-fit LL charge yield from Ref.~\cite{aalbers2024dark} interpreted in the TIB model framework.
}
\centering
\begin{tabular}{ cccccc } 
\hline
\hline
\noalign{\vskip 1mm}
Source & Energy [keV$_\text{ee}$] & $N_\text{i}$ & $N_\text{ex}/N_\text{i}$ & Drift field [V/cm] & $100 \times \frac{\alpha}{a^2 v_d}$ \\
\noalign{\vskip 0.5mm}
\colrule \noalign{\vskip 1mm}
\multirow{2}{*}{$\beta$} & \multirow{2}{*}{0.50--11.00} & \multirow{2}{*}{37--740} & \multirow{2}{*}{0.01--0.10} & 193  & 2.9 \\
                         &                              &                              &                               & 96.5 & 3.2 \\
\colrule \noalign{\vskip 1mm}
\multirow{2}{*}{M} & \multirow{2}{*}{1.07} & \multirow{2}{*}{79} & \multirow{2}{*}{0.01} & 193  & $3.5^{+0.3}_{-0.3}$\\
\noalign{\vskip 1mm}
                   &                       &                       &                        & 96.5 & 3.8$^{+0.8}_{-1.0}$\\
\noalign{\vskip 1mm}
\colrule \noalign{\vskip 1mm}
\multirow{2}{*}{L} & \multirow{2}{*}{5.19} & \multirow{2}{*}{370} & \multirow{2}{*}{0.05} & 193 & $4.3^{+0.2}_{-0.2}$\\
\noalign{\vskip 1mm}
                   &                       &                        &                        & 96.5 & $4.5^{+0.3}_{-0.3}$\\
\noalign{\vskip 1mm}
\colrule \noalign{\vskip 1mm}
MM & 2.05 & 150 & 0.02 & - & - \\
\colrule \noalign{\vskip 1mm}
LM & 6.01 & 420 & 0.06 & - & - \\
\colrule \noalign{\vskip 1mm}
LL & 10.04 & 680 & 0.10 & 96.5 & 5.1$^{+0.5}_{-0.5}$\\
\noalign{\vskip 1mm}
\hline
\hline
\end{tabular}
\label{tab:TIB_results}
\end{table*}
\addtolength{\tabcolsep}{-8pt}    

The TIB framework can then be applied to calculate a charge yield for LL decays, which was found to be $Q_y^\text{LL}/Q_y^\beta = 0.70 \pm 0.04$~\cite{aalbers2024dark} -- the priors used in the fit for that measurement are described in the Appendix.
Assuming the atomic emissions of LL-capture decays are the same as those in L-capture decays, such that the spatial extent of the ionization tracks is the same in both types of decay, we set the length scale $a_\text{LL}$ equal to $a_\text{L}$. 
The LL-decay has 1.94 times the energy of the L-shell, so its recombination can be estimated by scaling $N_\text{i}$ in Eq.~\ref{eq:xi}. 
A precise calculation taking into account the energy dependence of $N_\text{ex}/N_\text{i}$ gives $Q_y^\text{LL}/Q_y^\beta = 0.75 \pm 0.04$, in agreement with the best-fit value of $0.70\pm0.04$.
The uncertainty in the calculation is derived from the error on the MS-measured L-capture charge yield, which is used to determine $\alpha / (a_\text{L}^2v_d)$ (see Table~\ref{tab:TIB_results}).
The TIB model corresponding to $a_\text{L}$ and the best-fit LL charge yield are shown in Fig.~\ref{fig:Qy_E}. 

A similar calculation can be done for the MM-capture decay using the M-capture measurements, yielding $Q_y^\text{MM}/Q_y^\beta = 0.98 \pm 0.07$.
In this case the predicted value is higher than the measured $Q_y^\text{M}/Q_y^\beta = 0.91 \pm 0.04$, although the discrepancy between the WS2024 $\beta$ model and the original TIB fit at the M shell is also more significant. 
For LM decays, the contribution of the 1.1~keV M-shell component is subdominant in the ionization density of the 6.01~keV LM decay.
We calculate its recombination using $\alpha / (a_\text{L}^2v_d)$, and $N_\text{i}$ evaluated at the LM energy, giving $Q_y^\text{LM}/Q_y^\beta = 0.85 \pm 0.04$. 

\subsection{\label{subsubsec:impact}Impact on dark matter searches}

To assess the impact of DEC backgrounds on dark matter searches, we use a two-sided unbinned profile likelihood ratio test statistic in \{S1$c$, $\log_{10}$(S2$c$)\} to obtain the median WIMP sensitivity and discovery potential for a 1000 live-day projected exposure using the 5.5~tonne fiducial mass of LZ. 
Unless specified otherwise, the background components involved in these tests are modeled to match the best-fit values obtained in Ref.~\cite{aalbers2024dark}. 

\begin{figure}
    \includegraphics[width=1.0\linewidth]{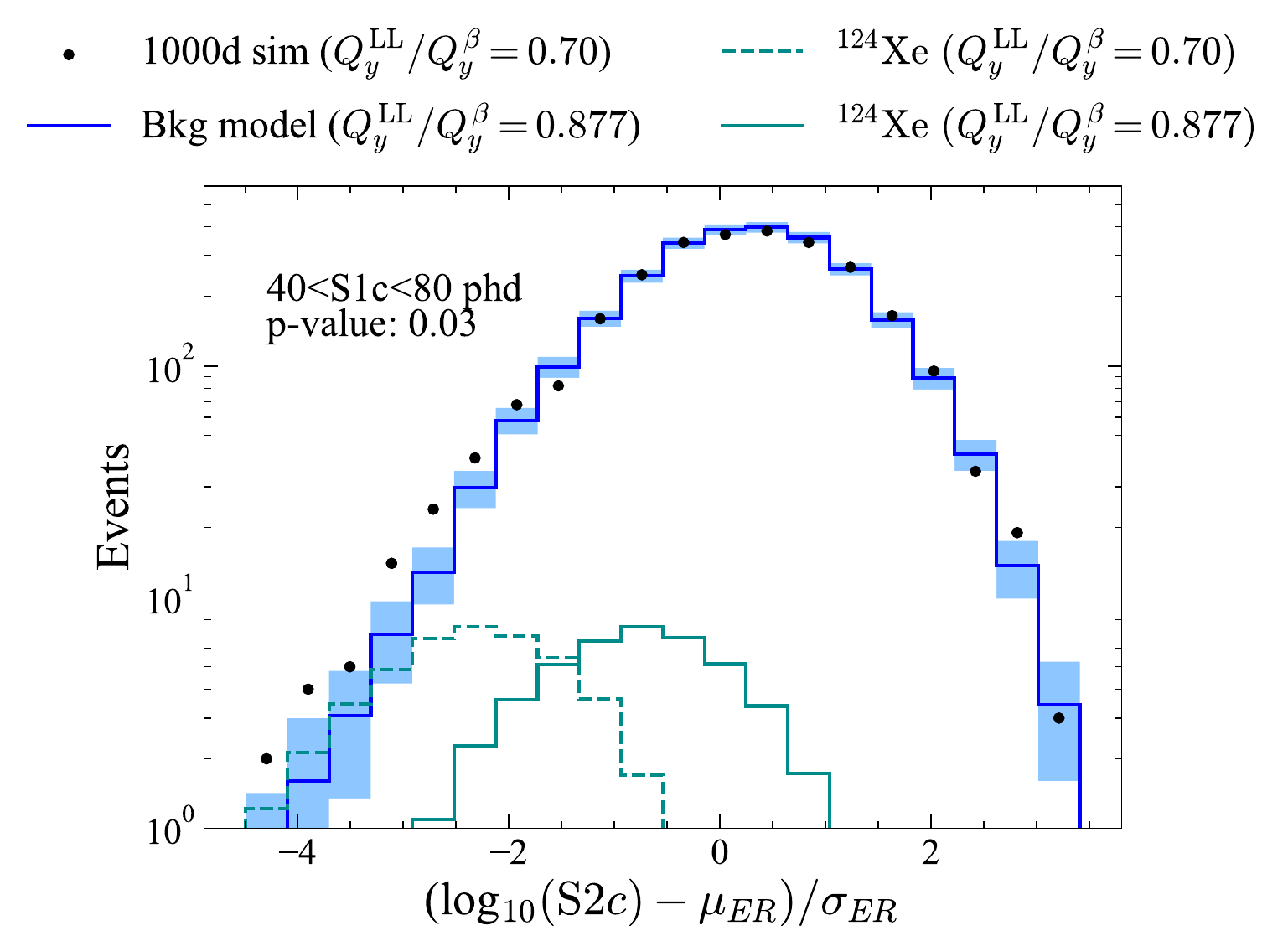}
    \caption{\justifying \label{fig:gof} 
    \small Simulated S2$c$ distribution, normalized to the mean and width of the background model as it varies with S1$c$, for a simulated 1000 live-day exposure of LZ with a fiducial mass of 5.5 tonnes (black points), drawn from the background model in Ref.~\cite{aalbers2024dark} with the best-fit $Q_y^\text{LL} / Q_y^\beta= 0.70$. 
    Only events with 40~phd $<$ ~S1$c$ $<$ 80~phd are shown, similar to the bottom panel of Fig. 4 in Ref.~\cite{aalbers2024dark}.
    A model using the measured $Q_y^\text{L}/Q_y^\beta = 0.877$ for the $^{124}$Xe LL component is fit to this dataset and overlaid in blue, with the $^{124}$Xe contribution alone in solid green. The more realistic ($Q_y^{LL} / Q_y^\beta= 0.70$) $^{124}$Xe distribution is represented by the dashed green bins.
    The blue shaded band depicts the central interval containing 68\% of the model's combined systematic and statistical uncertainties. The p-value for this fit is 0.03, showing that, at this exposure, incorrect modeling of the $^{124}$Xe component leads to significant tension with background-only data.
    }
\end{figure}

\begin{figure}
    \includegraphics[clip,trim={0.07cm 0.07cm 0.07cm 0.07cm},width=0.9\linewidth]{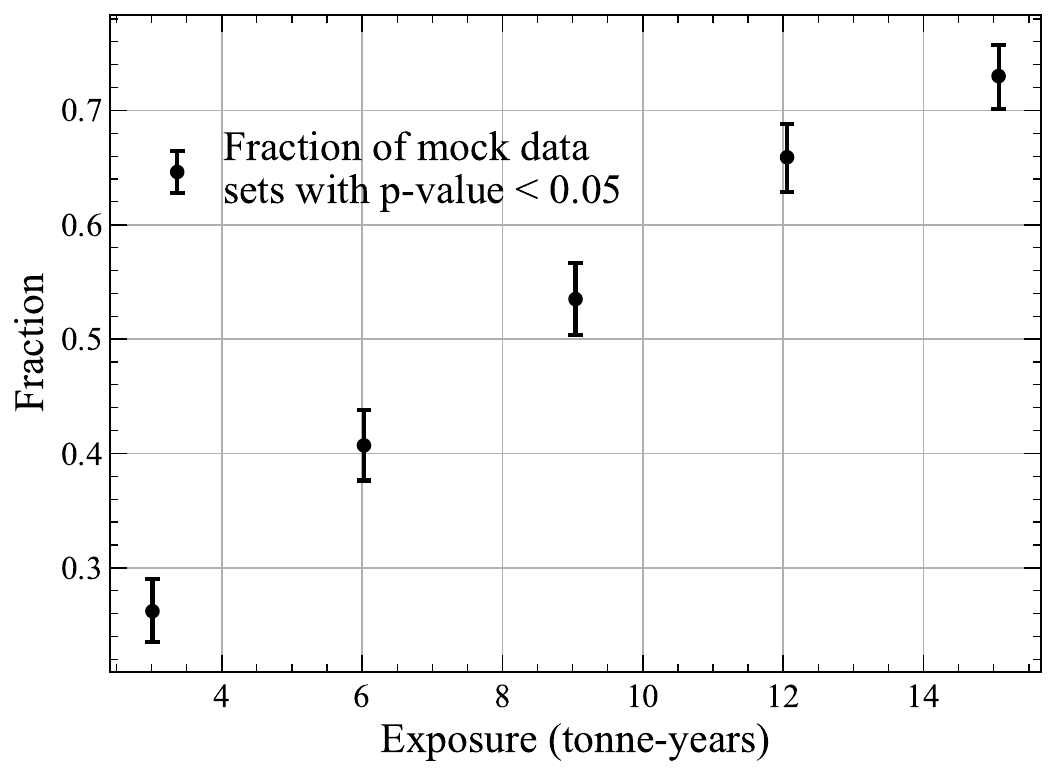}
    \caption{\justifying \small Average fraction of 1000 mock datasets at various exposures with failing p-values when compared to models without the enhanced recombination of LL decays. The maximum exposure shown here corresponds to a 1000~live day run with a 5.5~tonne fiducial mass.
    The p-value is calculated by comparing the S2$c$ distance to the ER median between mock data and model, as in the example of Fig.~\ref{fig:gof}.}
    \label{fig:p_values}
\end{figure}

We first emulate a scenario with no knowledge of the additional recombination reported in Ref.~\cite{aalbers2024dark}.
Mock data are simulated with a high-recombination LL ($Q_y^\text{LL}/Q_y^\beta = 0.70$) component. 
We then calculate the WIMP sensitivity 
and discovery potential 
when the model used to fit the mock data has an LL component with a charge yield ratio set at the L-capture measurement ($Q_y^\text{LL}/Q_y^\beta = 0.877$). 
Recombination fluctuations for the DEC events are generated as in Ref.~\cite{aalbers2024dark}, using the empirical resolution model in NEST~\cite{szydagis2review}.
For WIMPs heavier than 100~GeV/$c^2$, we observe a median sensitivity that is weaker by $\sim$10\% if the additionally suppressed LL charge yield is not taken into account, and the impact on the discovery potential is similar in scale. The relatively small size of this shift is likely due to the limited overlap of the $^{124}$Xe and WIMP spectra. Importantly, a goodness-of-fit test reveals significant tension between the background model and simulated data, which would call into question any analysis \RevA{of a 1000 live-day run of LZ} that did not adequately account for DEC charge yields. 
A histogram of the distance between observed S2$c$ and the median of the ER distribution for the mock dataset is provided in Fig.~\ref{fig:gof}.
The p-value comparing the mock data to the model without the additional LL charge suppression is 0.03, demonstrating that the nominal $^{124}$Xe charge yield is incompatible with an L-capture response model, much less a $\beta$-like one.
The fraction of mock datasets failing goodness-of-fit tests ($p<0.05$) for various exposures is shown in Fig.~\ref{fig:p_values}.

Next, we test the WIMP sensitivity for various charge yields and branching ratios of the MM, LN, LM, and LL decays. 
In these cases the mock data and the fitting templates are modeled identically.
For a baseline case we assume the following: 
\begin{enumerate}[I.]
    \item The MM decay (branching ratio 0.13\%) has a charge yield ratio $Q_y^\text{MM}/Q_y^\beta$ equal to the WS2024 MS-measured value of $Q_y^\text{M}/Q_y^\beta = 0.913$.
    \item Both LM (0.49\%) and LN (0.27\%) decays have charge yield ratios $Q_y^\text{LM}/Q_y^\beta$ and $Q_y^\text{LN}/Q_y^\beta$ equal to the WS2024 MS-measured value of $Q_y^\text{L}/Q_y^\beta = 0.877$.
    \item The LL decay (1.22\%) has a charge yield ratio $Q_y^\text{LL}/Q_y^\beta = 0.70$ as reported in Ref.~\cite{aalbers2024dark}.
\end{enumerate}
The DEC charge yields are individually varied to the conservatively low values in the Appendix, and the branching ratios (see also Table~\ref{tab:dec_counts}) are varied by $\pm40$\%.
Again, compared to the baseline model, the worst cases result in a $\sim$10\% weaker median sensitivity for 30--300~GeV/$c^2$ WIMPs in the high-recombination MM case, and for WIMPs heavier than 50~GeV/$c^2$ in the high-recombination LL case. 

Overall, the shift in charge yield due to increased ionization density does not impact the ultimate dark matter sensitivity of LZ by more than $\sim$10\%, primarily because of the limited overlap between the signal distributions and DEC peaks for WIMP masses below $O(100)$~GeV/$c^2$. 
However, dark matter candidates with signal spectra that span higher energies up to the 64.3~keV KK-capture peak, such as some effective field theory models~\cite{aalbers2024first}, are likely to be more significantly affected.

\section{\label{sec:discussion}Conclusion}

M- and L-shell electron-capture events in LZ are observed to have lower charge yields than equivalent-energy $\beta$ decays, confirming earlier measurements~\cite{akerib2017ultralow,temples2021measurement}. 
The TIB model can qualitatively explain the charge yield suppression if the track ionization density correlates with the amount of recombination, but the model cannot predict the size of the effect \emph{a priori}.
Across the 193~V/cm and 96.5~V/cm LZ runs, as well as the LUX and XELDA results up to 363~V/cm, there appears to be no conclusive field dependence for the size of the recombination enhancement as quantified by the charge yield ratios.

The recombination profile of K-capture events across both runs is more ambiguous.
K-capture events in the 193~V/cm run have lower charge yields than the multi-site expectation of the $\gamma + \beta$ (X-ray + {\color{black} Auger-Meitner}) models. 
%In the 96.5~V/cm run, where the MS and SS measurements are in tension, the measurement is consistent with the $\gamma + \beta$ expectation.
In the 96.5~V/cm run the measurement is consistent with the $\gamma + \beta$ expectation.
We emphasize that the $\gamma$ model is not yet validated with LZ data. 

The recombination of $^{124}$Xe LL events is enhanced even more than in EC events; the fitted LL charge yield in the WS2024 dark matter search is only 70\% of the equivalent-energy $\beta$ yield~\cite{aalbers2024dark} while the M- and L-capture charge yields presented in this work are around 90\% of their respective equivalent-energy $\beta$ yields.
We use the TIB model to calculate the LL charge yield from the L-capture measurement by assuming that the volume in which recombination happens is the same for both decays. 
This calculation agrees with the fitted LL charge yield in the WS2024 dark matter search~\cite{aalbers2024dark}.

For the first time, we have reached exposures that require the precise modeling of significant recombination in extremely rare decays of $^{124}$Xe, which now forms a crucial component of the background model for current and future WIMP searches.
In our current assessment, if DEC charge yields are accounted for properly, the impact on future WIMP searches is less than $\sim$10\% for the scenarios tested here. Because WIMP elastic scattering signatures are relatively free of leakage from DEC events, \RevA{inferred WIMP constraints} are not significantly impacted even without proper treatment of DEC charge suppression. However, \RevA{at least for the drift fields explored here}, a background model without these effects would not accurately describe either current or future WIMP search datasets. 
Upcoming exposures refining the DEC charge yields and characterizing the low-charge tails are required to confirm this assessment.

\subsection{\label{sec:future}Future work}

The data in this work and Refs.~\cite{akerib2017ultralow, temples2021measurement, aalbers2024dark} should be combined and extended to characterize EC and DEC charge yield
distributions, as well as to extract trends with drift field, energy, and ionization density.
Since the MS selection provides a charge-only measurement, and the SS selection is limited by low statistics, efforts to develop novel calibration sources such as $^{131}$Cs, which decays via EC directly to the nuclear ground state (i.e.~without an associated $\gamma$ ray), should be pursued to obtain large samples of SS EC events in the LXe bulk.

The charge yield of $^{124}$Xe LL events could only be assessed in a fit to the final WS2024 data set, with direct charge yield measurements of $^{124}$Xe DECs that overlap with WIMP spectra limited by low branching ratios. Nevertheless, atomic de-excitations of the more frequent KK- and KL-capture decays can proceed through LL and LM vacancy states via X-ray emission, enabling the extraction of the LL and LM charge yields.
Other isotopes that create double vacancies, such as an EC decay with a low-energy IC electron, could also provide insights into the recombination of true DEC events. 

The qualitative aspects of the TIB model paradigm are improved by including length scales beyond the single box size.
With just two characteristic length scales, for instance -- one for the electron range and another for the ionization track -- the recombination of ER events up to 200~keV$_\text{ee}$ can be described~\cite{Dahl:2009nta}. 
A measure of overlap defined on the electron range and ionization track scales might be able to replace the role of ionization density in recombination, which lacks a first-principles description. 
In this case, the track structure of EC, DEC, $\beta$, photoabsorption, and Compton scatters would be keys to predict their distinctive yields~\cite{obodovskii1994scintillation, aprile2012measurement, baudis2013response, szydagis2021investigating}.
Computational models of recombination that fully account for track structure, electron transport, and other dynamics in $N_\text{i} \approx 10^3$ systems (covering WIMP searches) could be within reach. 

The enhanced recombination observed in EC and DEC decays leads to conjectures about three types of observable events in xenon TPCs:
\begin{enumerate}
    \item In the next generation of xenon TPC dark matter experiments, neutrino-electron scattering will be the dominant source of ER background~\cite{aalbers2022next}, with some probability to interact with inner shell electrons.
    These neutrino-induced ISVs have more recombination than EC events due to the overlapping tracks from the {\color{black} Auger-Meitner} cascade and the ejected electron.
    \item Single $\beta$ decays and Compton scattering $\gamma$ rays are backgrounds in the search for the neutrinoless double $\beta$ decay of $^{136}$Xe~\cite{akerib2020projected}.
    Each background component and the signal have different recombination fractions due to varying degrees of track overlap.
    \item The Migdal effect is a hypothesized ER signal that accompanies an NR event, which may extend the reach of xenon TPCs to sub-GeV/$c^2$ WIMPs~\cite{xu2024search}.
    Migdal events will have additional recombination due to the overlap of NR and ER tracks. 
\end{enumerate}

\begin{acknowledgments}
The LZ Collaboration acknowledges the key contributions of Dr. Sidney Cahn, Yale University, in the production of calibration sources.
The authors also thank Areej Al Musalhi for her assistance with the design of illustrative figures.
We respectfully acknowledge that we are on the traditional land of Indigenous American peoples and honor their rich cultural heritage and enduring contributions.
The research supporting this work took place in part at the Sanford Underground Research Facility (SURF) in Lead, South Dakota. Funding for this work is supported by the U.S. Department of Energy, Office of Science, Office of High Energy Physics under Contract Numbers DE-AC02-05CH11231, DE-SC0020216, DE-SC0012704, DE-SC0010010, DE-AC02-07CH11359, DE-SC0015910, DE-SC0014223, DE-SC0010813, DE-SC0009999, DE-NA0003180, DE-SC0011702, DE-SC0010072, DE-SC0006605, DE-SC0008475, DE-SC0019193, DE-FG02-10ER46709, UW PRJ82AJ, DE-SC0013542, DE-AC02-76SF00515, DE-SC0018982, DE-SC0019066, DE-SC0015535, DE-SC0019319, DE-SC0024225, DE-SC0024114, DE-AC52-07NA27344, \& DE-SC0012447. This research was also supported by the U.S. National Science Foundation; the UKRI’s Science \& Technology Facilities Council under award numbers ST/W000490/1, ST/W000482/1, ST/W000636/1, ST/W000466/1, ST/W000628/1, ST/W000555/1, ST/W000547/1, ST/W00058X/1, ST/X508263/1, ST/V506862/1, ST/X508561/1, ST/V507040/1 , ST/W507787/1, ST/R003181/1, ST/R003181/2,  ST/W507957/1, ST/X005984/1, and ST/X006050/1; the Portuguese Foundation for Science and Technology (FCT) under award numbers PTDC/FIS-PAR/2831/2020; the Institute for Basic Science, Korea (budget number IBS-R016-D1); and the Swiss National Science Foundation under award number 10001549. This research was supported by the Australian Government through the Australian Research Council Centre of Excellence for Dark Matter Particle Physics under award number CE200100008. We acknowledge additional support from the UK Science \& Technology Facilities Council (STFC) for PhD studentships and the STFC Boulby Underground Laboratory in the United Kingdom, the GridPP~\cite{faulkner2005gridpp, britton2009gridpp} and IRIS Collaborations, in particular at Imperial College London and additional support by the University College London Cosmoparticle Initiative, and the University of Zurich. We acknowledge additional support from the Center for the Fundamental Physics of the Universe, Brown University. K.T.L acknowledges the support of Brasenose College and Oxford University.  This research used resources of the National Energy Research Scientific Computing Center, a DOE Office of Science User Facility supported by the Office of Science of the DOE under Contract No. DE-AC02-05CH11231. We gratefully acknowledge support from GitLab through its GitLab for Education Program. The University of Edinburgh is a charitable body, registered in Scotland, with the registration number SC005336. The assistance of SURF and its personnel in providing physical access and general logistical and technical support is acknowledged. We acknowledge the South Dakota Governor's office, the South Dakota Community Foundation, the South Dakota State University Foundation, and the University of South Dakota Foundation for the use of xenon. We also acknowledge the University of Alabama for providing xenon. 
\end{acknowledgments}

\appendix

\vspace{5mm}
\section{\label{app:scaling_argument}DEC modeling in WS2024}
For completeness, we describe the initial calculation used to provide conservative bounding cases for the DEC charge yield in LZ's recent WIMP search~\cite{aalbers2024dark}.
Only LM and LL DEC decays were included in the background model as information on other shells was not available at the time. 
The ratio $Q_y^\text{LM}/Q_y^\beta$ was set at the $Q_y^\text{L}/Q_y^\beta$ measurement, while $Q_y^\text{LL}/Q_y^\beta$ was allowed to float within a prescribed range. 
The upper bound of the range was taken to match $Q_y^\text{L}/Q_y^\beta$, under the expectation that LL events would have at least as much charge yield suppression as the single EC. 
A value $\xi_\text{LL}^\text{high}$ that matches the upper bound was calculated such that 
\begin{equation}\label{eq:LL_starting}
    \frac{Q_y^\text{LL} (\xi_\text{LL}^\text{high})}{Q_y^\beta (10.04~\text{keV})} = \frac{Q_y^\text{L}}{Q_y^\beta (5.1881~\text{keV})}.
\end{equation}

We used the TIB model to derive a conservative lower bound for the LL charge yield, given by $\xi_\text{LL}^\text{low}$, treating the effective ionization density as the driving variable. 
We obtain $\xi_\text{LL}^\text{low}$ by shrinking the box size until the effective ionization density is doubled relative to the starting density at $\xi_\text{LL}^\text{high}$. 
Since $\xi$ in Eq.~\ref{eq:xi} is proportional to $N_\text{i}/a^2$ instead of $N_\text{i}/(8a^3)$, doubling the density increases $\xi$ by a factor of $2^{2/3}$ (instead of 2) such that $\xi_\text{LL}^\text{lower} = 1.59 \xi_\text{LL}^\text{high}$.
Table~\ref{tab:xi_fit} shows the corresponding ranges of $4\xi/N_\text{i} = \alpha/(a^2 v_d)$ for the relevant DEC decays.

\begin{figure}[!t]
    \centering
    \includegraphics[width=0.95\linewidth]{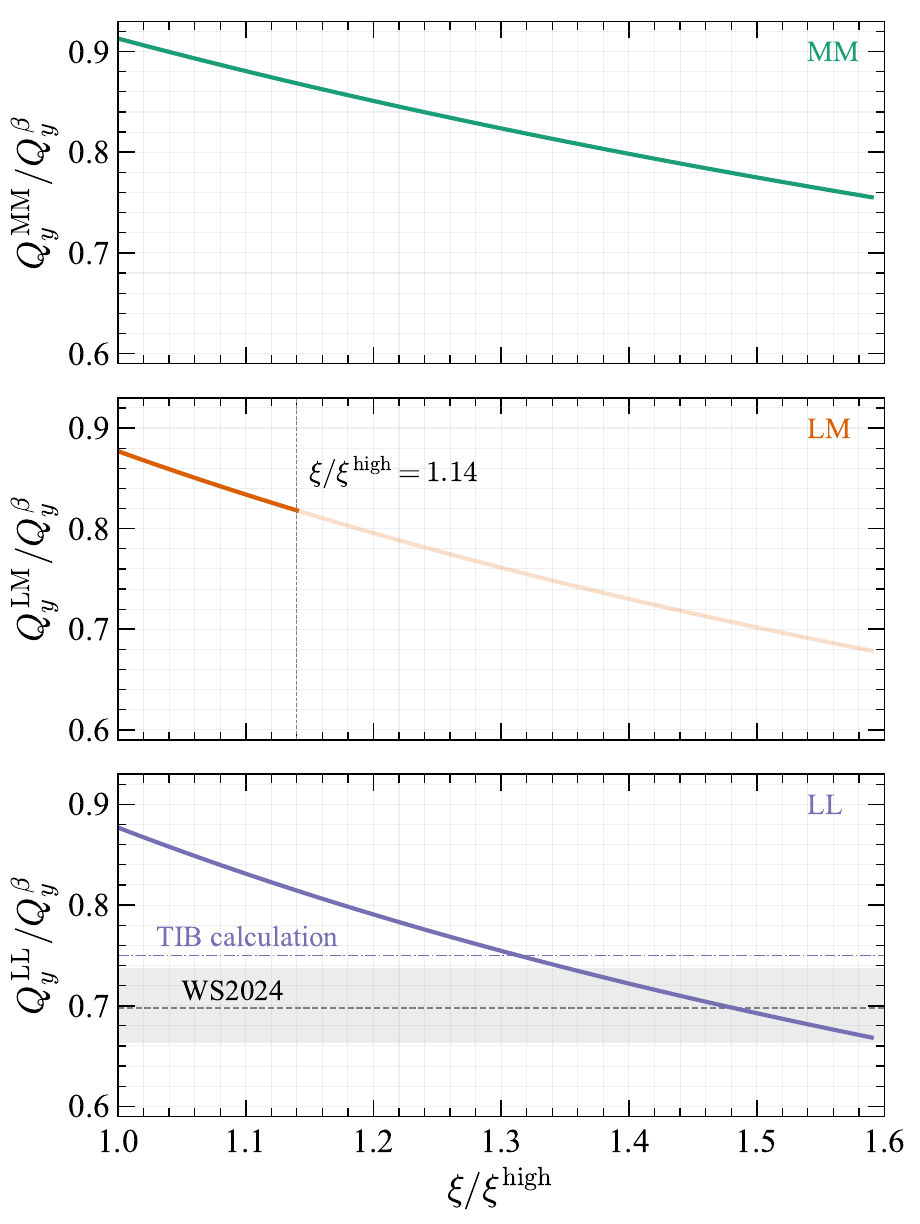}
    \caption{\justifying \small Charge yields of the MM (top), LM (middle), and LL (bottom) decays as a function of the scaling factor of $\xi$, expressed as ratios of the $\beta$ decay yields at $2.05$~keV, $6.01$~keV, and $10.04$~keV, respectively. 
    The span of $\xi/\xi^\text{high}$ reflects the increase in ionization density. 
    For the MM and LL decays, $\xi/\xi^\text{high}$ spans from $1$ (no increase) to $1.59$ (doubled ionization density), whereas for the LM decays $\xi/\xi^\text{high}$ attains a maximum of $1.14$ (an ionization density increase of 20\%). 
    %Doubling $\rho_N$ (shrinking the box) increases $\xi$ by a factor of 1.59 instead of 2 because $\xi$ is proportional to $N_\text{i}/a^2$ instead of to $\rho_N$.
    Also shown in the bottom panel is the best-fit LL charge yield (black dashed) from Ref~\cite{aalbers2024dark}, and the TIB calculation (purple dot-dashed) of the effect from Sec.~\ref{subsec:box_model}.
    }
    \label{fig:qyRatio_vs_box}
\end{figure}

This is a conservative range for LL decays because the starting point ($\xi_\text{LL}^\text{high}$ given in Eq.~\ref{eq:LL_starting}) already has a higher ionization density relative to L-capture events. 
Letting $\rho_\text{L}$ be the ionization density of L-capture events, we find that $\rho^\text{high}_\text{LL} = 1.23 \rho_\text{L}$, using the MS-measured L-capture ratio of 0.877.
The lower bound on the LL charge yield is obtained by doubling $\rho_\text{LL}^\text{high}$, such that $\rho_\text{LL}^\text{low} = 2\rho_\text{LL}^\text{high} = 2.5\rho_\text{L}$, a larger density than predicted by the calculation in Sec.~\ref{subsec:box_model}.
The profiles in Fig.~\ref{fig:qyRatio_vs_box} visualize how the modeled charge yield for DEC events decreases with increasing $\xi$. 
The WS2024 best-fit ratio $Q_y^\text{LL}/Q_y^\beta = 0.70 \pm 0.04$ corresponds to an ionization density of $(2.2 \pm 0.3)\rho_\text{L}$.

\begin{table}[h!]
\caption{\justifying \small TIB model parameters corresponding to the upper and lower bounds of the range described in the text, for the 96.5~V/cm run. 
The smaller number in the range sets the LM and LL charge yield ratios to match the MS-measured L-capture ratio, and the MM charge yield ratio to match the MS-measured M-capture ratio. 
The larger number corresponds to a doubled effective ionization density.
TIB parameters for the corresponding EC decays are also shown (from Table~\ref{tab:TIB_results}).}
\begin{ruledtabular}
\begin{tabular}{l c c c c} 
Source & $N_\text{i}$ & $N_\text{ex}/N_\text{i} $ & $100 \times \frac{\alpha}{a^2v_d}$ range & $100 \times \frac{\alpha}{a^2v_d}$ (EC) \\ [0.5ex] 
\colrule \noalign{\vskip 1mm}
 MM & 150 & 0.02 & 4.7--7.5 & 3.8$^{+0.8}_{-1.0}$ \\
 \noalign{\vskip 0.5mm}
\colrule 
 \noalign{\vskip 0.5mm}
 LM & 420 & 0.06 & 4.3--4.9 & \multirow{2}{*}{4.5$^{+0.3}_{-0.3}$} \\
 LL & 680 & 0.10 & 3.5--5.5 & \\
\end{tabular}
\end{ruledtabular}
\label{tab:xi_fit}
\end{table}

\clearpage
\bibliographystyle{apsrev4-2}
\bibliography{main}

\end{document}